\documentclass[apj]{emulateapj}
\usepackage{multirow}
\usepackage{booktabs}
\usepackage{graphicx}
\usepackage{amsmath}
\usepackage{longtable}
\usepackage{rotating}

\newcommand{\ts}{t_{\rm s}}
\newcommand{\Ts}{T_{\rm s}}
\newcommand{\sigmag}{\Sigma_{\rm g}}
\newcommand{\sigmasg}{\Sigma_{\rm sg}}
\newcommand{\sigmabg}{\Sigma_{\rm bg}}
\newcommand{\deltagap}{\Delta_{\rm gap}}
\newcommand{\hg}{h_{\rm g}}
\newcommand{\hsg}{h_{\rm sg}}
\newcommand{\hbg}{h_{\rm bg}}
\newcommand{\msun}{M_\odot}
\newcommand{\mj}{M_{\rm J}}
\newcommand{\mplanet}{M_{\rm p}}
\newcommand{\modelzero}{1\times0.2M_{\rm J}}
\newcommand{\modelone}{1\times1M_{\rm J}}
\newcommand{\modeltwo}{4\times1M_{\rm J}}
\newcommand{\modelthree}{4\times2M_{\rm J}}
\newcommand{\modelhltau}{3\times0.2M_{\rm J}}
\newcommand{\icarus}{{\it Icarus}}

\begin{document}
\title{Observational Signatures of Planets in Protoplanetary Disks I: Gaps Opened by Single and Multiple Young Planets in Disks}

\shorttitle{Gaps Opened by Multiple Planets in Protoplanetary Disks}

\shortauthors{Dong, Zhu, \& Whitney}

\author{Ruobing Dong\altaffilmark{1,2,5}, Zhaohuan Zhu\altaffilmark{3,5}, Barbara Whitney\altaffilmark{4}}

\altaffiltext{1}{Nuclear Science Division, Lawrence Berkeley National Lab, Berkeley, CA 94720, rdong2013@berkeley.edu}
\altaffiltext{2}{Department of Astronomy, University of California at Berkeley, Berkeley, CA 94720}
\altaffiltext{3}{Department of Astrophysical Sciences, Princeton University, Princeton, NJ 08544}
\altaffiltext{4}{Astronomy Department, University of Wisconsin-Madison, 475 N. Charter St., Madison, WI 53706, USA}
\altaffiltext{5}{Hubble Fellow}

\clearpage

\begin{abstract}

It has been suggested that the gaps and cavities recently discovered in transitional disks are opened by planets. To explore this scenario, we combine two-dimensional two fluid (gas + particle) hydrodynamical calculations with three-dimensional Monte Carlo Radiative Transfer simulations, and study the observational signatures of gaps opened by one or several planets, making qualitative comparisons with observations. We find that a single planet as small as 0.2 $\mj$ can produce a deep gap at millimeter (mm) wavelengths and almost no features at near-infrared (NIR) wavelengths, while multiple planets can open up a few $\times$10 AU wide common gap at both wavelengths. Both the contrast ratio of the gaps and the wavelength dependence of the gap sizes are broadly consistent with data. We also confirm previous results that NIR gap sizes may be smaller than mm gap sizes due to dust-gas coupling and radiative transfer effects. When viewed at a moderate inclination angle, a physically circular on-centered gap could appear to be off-centered from the star due to shadowing. Planet-induced spiral arms are more apparent at NIR than at mm wavelengths. Overall, our results suggest that the planet-opening-gap scenario is a promising way to explain the origin of the transitional disks. Finally, inspired by the recent ALMA release of the image of the HL Tau disk, we show that multiple narrow gaps, well separated by bright rings, can be opened by 0.2$\mj$ planets soon after their formation in a relatively massive disk.


\end{abstract}



\section{Introduction}\label{sec:intro}

Flattened, rotating gaseous protoplanetary disks around young stars reprocess light from the central star, modifying the spectral energy distribution (SED) of the system, and revealing themselves in resolved images at various wavelengths \citep{williams11}. These disks are considered to be the birth place of planets \citep{armitage11}. Ideally, detecting newly-born planets in protoplanetary disks directly is one of the best ways to constrain planet formation, as it can reveal {\it when}, {\it where}, and {\it how} do planets form. 
However, it is difficult to directly detect forming planets in disks, and only a few have been identified so far \citep[e.g.][]{huelamo11, kraus12, quanz13-planet, brittain14, biller12, close14, reggiani14}. Therefore, methods for indirect detection of young planets in disks are necessitated.

In a circumstellar disk, planets excite asymmetric structures such as spiral density waves, and may clear material around their orbits to form gaps, through gravitational disk-planet interactions \citep[e.g.][]{goldreich80, lin93, bryden99, kley12}. While directly detecting planets in disks is hard, these large scale planet-induced distortions are more prominent, and may be detectable in observations with high spatial resolution \citep[e.g.][]{wolf05, varniere06-gap, jang-condell07, jang-condell09, fouchet10, jang-condell12, gonzalez12, ruge13, dejuanovelar13}. By identifying and comparing these features with theoretical models of disk-planet interactions, we can learn much about the possibly embedded planets. 

The last few years have witnessed ground breaking results in resolved observations of protoplanetary disks, and fine structures possibly produced by planets have been found in many systems at multiple spectral windows. At near-infrared (NIR) wavelengths, more than a dozen nearby disks have been imaged by 10-meter class mirrors, in particular VLT (e.g. \citealt{quanz11, quanz12, canovas13, quanz13-gap, garufi13, garufi14, avenhaus14}) and Subaru (the Subaru Strategic Exploration of Exoplanets and Disks Survey, \citealt{tamura09, tsukagoshi14, takami13, follette13, grady13, tanii12, mayama12, kusakabe12, muto12, hashimoto11}).These observations took advantage of the Polarimetric Differential Imaging technique (PDI, e.g. \citealt{perrin04, hinkley09}) to effectively remove the unpolarized stellar light while retaining the polarized component in the scattered light from the dust grains in disks. Inner working angles (the smallest angular separation from the central source to which NIR observations have access) on the order of $0.1\arcsec$ and diffraction limited angular resolution $\sim0.04-0.06\arcsec$ at $J$, $H$, $K$~bands have been routinely achieved (corresponding to $\sim14$~AU and $\sim6-8$~AU at the distance of nearby star forming regions, $\sim140$~pc, such as Taurus).

Accompanying the progress in NIR direct imaging, radio interferometry has opened up another window lately around $\sim$1~mm for detailed disk structure studies. Spatially resolved observations of dust continuum and/or molecular line emission have been carried out for a few dozens nearby disks using the Submillimeter Array \citep[e.g.][]{brown09, andrews09, andrews10, andrews11}, the Combined Array for Research in Millimeter-wave Astronomy \citep[e.g.][]{isella09, isella10, ricci13}, the Plateau de Bure Interferometer \citep[e.g.][]{pietu06, guilloteau11}, and the newly commissioned Atacama Large Millimeter Array (ALMA) \citep[e.g.][]{vandermarel13, casassus13, fukagawa13, perez14, zhang14}. Protoplanetary disks are generally optically thick at NIR, so NIR imaging traces the surface structure \citep[e.g.][]{takami14}. On the other hand, they are often optically thin in mm dust continuum\footnote{In this work, we call observations at ALMA wavelengths, e.g. ~0.3-3 mm wavelengths ``mm observations''} and certain molecular line emissions, so mm observations can probe the distribution of material in the midplane regions of disks. As ALMA is transitioning into its full capacity phase in the next few years, it will provide more exciting results with its sub-$0.1\arcsec$ angular resolutions and superb sensitivity.

Among disks with possible planet-induced structures, a particularly interesting group is the so called transitional disks\footnote{In this work, we call both ``transitional'' and ``pretransitional'' disks as defined in \citet{espaillat14} transitional disks.}. Discovered through their unique infrared deficit from NIR to $\sim10~\micron$, which signals a lack of warm dust in the inner disk \citep{calvet05, espaillat07, espaillat10}, these disks have been subsequently shown to harbor large gaps or cavities with sizes often up to tens of AU in spatially resolved observations (e.g., \citealt{thalmann10, mayama12, hashimoto12, hughes09, andrews11, zhang14, perez14}). The appearances of transitional disks are not always consistent at different wavelengths. In some cases, NIR images (tracing small $\sim\micron$-sized grains) do not show the cavities seen at mm (tracing big $\sim$mm-sized grains)  down to their inner working angle (\citealt{dong12cavity}), while in some other cases the cavities in scattered light and/or gas observations appear to be smaller than in the dust continuum \citep[e.g.][]{garufi13, zhang14, perez14hd142527}. 

The formation of the gap/cavity in transitional disks is not well understood at the moment. Several mechanisms can partially explain the observations, including grain growth \citep[e.g.][]{dullemond05, birnstiel12}, photoevaporation (e.g. \citealt{alexander07, alexander09, owen12}, see also \citealt{rosotti13}), infall from counterrotating external environments \citep{vorobyov14}, and of particular interest, disk-planet interactions \citep{bryden99, varniere06-planet}. While the gap opened by a single planet may be too narrow to match observations, a system of several giant planets may open a combined/common gap with a size comparable to observed values. \citet{zhu11} and \citealt{dodsonrobinson11} performed two-dimensional (2D) hydro disk-planet simulations, and found that a system with 4 planets can indeed open a wide gap in the gas surface density and still maintain a moderate accretion rate onto the star as observed in some systems. More recently, dust particles have been added into these models. \citet{zhu12} and \citet{pinilla12b} have shown that the ``dust filtration'' effect \citep{paardekooper06, rice06} may allow the cavity in the gas and dust particles with different sizes to have different sizes (see also \citealt{pinilla12a}).  This is because the pressure maximum outside the gap created by a planet can efficiently pile up the big grains ($\sim$mm-sized) with a stopping time close to unity, and prevent them from entering the gap, while still allowing gas and smaller size grains (e.g. $\micron$-sized or smaller) to move into the gap. 


Observational signatures of gaps opened by planets has been explored in the past. \citet{fouchet10} and \citet{gonzalez12} carried out 3D Smoothed Particle Hydrodynamics (SPH) gas+dust calculations and MCRT simulations to make predictions for ALMA observations of planet-induced gaps. In a series of studies by \citet{pinilla12b}, \citet{dejuanovelar13}, and \citet{pinilla14}, the authors ran hydro simulations to calculate 2D gas surface density, and fed it into a 1D dust coagulation/fragmentation models to calculate the size evolution and spatial distribution of dust in disks. Synthetic observations with Very Large Telescope (VLT)/SPHERE-ZIMPOL, Subaru/HiCIAO, and ALMA were produced with self-consistent 2D (radial-polar) MCRT simulations. \citet{pinilla12b} and \citet{dejuanovelar13} pointed out that dust filtration could pile up big grains at the edge of the gas gap opened by one giant planet to form a ring at mm wavelengths, while allowing small grains to enter the gap to produce scattered light. \citet{pinilla14} expanded the work to have two planets at a large separation, which opened two non-overlapping gaps. 


This paper is the first in a series in which we explore various observational signatures of planets in protoplanetary disks. We intend to bridge the gap between theory and observation, by combining tools of hydro/MHD calculations with Monte Carlo Radiative Transfer (MCRT) simulations. Specifically, in order to directly translate hydro/MHD simulations into model observations, we use the \citet{whitney13} code to read in an external numerical density grid and calculate the radiative transfer. In this study, we focus on gaps opened by one and multiple planets, and aim at answering this basic question: are they broadly consistent with some observed gaps in disks? Since most observations only probe dust in disks, we need to consider dust dynamics separately from gas dynamics to make fair comparisons with these observations. In this work, we expand previous studies by combining 2D (radial and azimuthal) two fluid (gas + particle) hydro calculations with 3D MCRT simulations (the 3rd dimension is the height of the disk that we calculate prior to reading into the MCRT code). Images at both NIR and mm wavelengths are produced, and detailed comparisons with observations are made. The paper is organized as follows: In Section~\ref{sec:setup} we introduce our hydro and MCRT methods. The main results are presented in Section~\ref{sec:results}. We discuss our results in Section~\ref{sec:discussion}, followed by a short summary in Section~\ref{sec:summary}.




\section{Simulation Setup}\label{sec:setup}

The general workflow to produce model observations is as follows:  
\begin{enumerate}
\item We carry out global 2D (radial-azimuthal) two-fluid (gas and dust particles) hydrodynamical simulations to calculate the surface density distribution of the gas and 1~mm dust particles (big dust) in disks with one or more planets. At the beginning of the hydro simulations, the surface density of the big grains is $0.9\%$ of the gas. After the hydro simulations, we then linearly convert the surface density of the gas to the surface density of the small dust ($\micron$-sized or smaller) assuming a 1000:1 mass ratio, as the small dust is always well coupled to the gas in our models.
\item The resulting 2D disks of both dust populations are puffed up into 3D structures assuming Gaussian density profiles in the vertical direction. The resulting 3D disk structures are fed into MCRT simulators, which then produce the corresponding SED and raw model images at both $H$ band and ALMA band 7.
\item The raw images at both wavelengths are convolved by Gaussian point spread functions (PSF) to produce convolved images that have angular resolutions comparable to observations.
\end{enumerate}

\subsection{Hydrodynamical Gas+Dust Simulations}\label{sec:hydro}
We have carried out global 2D (radial-azimuthal) two-fluid (gas and dust particles) hydrodynamical simulations
using the FARGO code \citep{masset00} with a newly implemented dust fluid \citep{zhu12}. 
The dust is treated as a zero pressure fluid and couples with the gas via drag terms. 
No feedback from the dust on the gas is considered since gas-to-dust mass ratio is always
much larger than 1 in our simulations.  The drag terms are computed by assuming
the dust is in the Epstein regime \citep{whipple72, weidenschilling77}, 
which is always true for our adopted particle size (1~mm) and disk parameters. 
At the disk midplane, the dust
stopping time $\ts$ in the Epstein regime can be written as \citep{zhu12}
\begin{equation}
\ts=\frac{\pi s \rho_{d} }{2\sigmag\Omega},
\label{eq:ts}
\end{equation}
where $s$ is the size of the particle, $\rho_{d}$ is the density of the particle (assumed to be 1 g cm$^{-3}$), $\sigmag$ is the surface density of the gas, and $\Omega$ is the angular velocity of the disk. When the dust's stopping time is much shorter than the disk's dynamical timescale (always true for the 1~mm particles; $\ts$ is $\lesssim10\%$ of the orbital period), we can use  the ÒShort Friction
Time ApproximationÓ(SFT, \citealt{johansen05}) to calculate the dust velocity: 
\begin{equation}
\bold{v_{d}}=\bold{v_{g}}+\ts\frac{\nabla P}{\sigmag},
\label{eq:eqSFT}
\end{equation}
where $\bold{v_{d}}$ is the velocity of the dust particle, $\bold{v_{g}}$ is the velocity of the gas, and $P$ is the gas pressure. 
This approximation is  always true for our chosen particle size (more discussion and the detailed comparison with other approaches are given in the Appendix of \citealt{zhu12}).
Dust turbulent diffusion is modeled as a diffusion
term in the dust continuity equation. The Schmidt number $Sc$, which is
defined as the ratio between the total accretion stress 
and particle mass diffusivity, is assumed to be 1 \citep{johansen05}. Dust settling effect is taken into account when 2D disk structures are puffed up into 3D structures in Section~\ref{sec:mcrt}.

Hydrodynamical model setups are largely adopted from \citet{zhu11}, and are briefly summarized here.
We assume a central stellar mass of $1 \msun$ and a fully viscous disk.
We further assume a radial temperature distribution
$ T = 221 (r/{\rm AU})^{-1/2}$~K, which is roughly consistent with typical T Tauri
disks in which irradiation from the central star dominates the disk temperature
distribution \citep[e.g.][]{dalessio01}. The disk is vertically isothermal, and the gas scale height $\hg$ is $\hg=c_{\rm s}/\Omega$, where $c_{\rm s}$ is the sound speed.
The adopted radial temperature distribution corresponds to $\hg/r = 0.029 (r/{\rm AU})^{0.25}$ assuming vertical hydrostatic equilibrium.
We set $\alpha = 0.001$. With this $\alpha$, the gap edge will develop a vortex only if massive planets are in the disk and the vortex cannot quickly dissipate \citep{zhu14stone, fu14}. This is consistent with the result that no vortex is observed at the end of all current simulations.  The initial gas surface density is
\begin{equation}
\sigmag=178\frac{\rm AU}{r}e^{-\frac{r}{\rm 100AU}}\ {\rm g\ cm^{-2}},
\label{eq:sigmags}
\end{equation}
from $r \sim 1 - 500$~AU, so that it reaches a steady
disk solution with an accretion rate $\dot{M} \sim 10^{-9} M_\odot$ yr$^{-1}$,
typical of T Tauri disks \citep{gullbring98, hartmann98}.

The dust particles in our hydro runs are assumed to be 1 mm in radius, so they represent the ``big grains'' in disks that are mainly responsible for mm dust continuum emission. The surface density of the particles (big grains) $\sigmabg$ is set to be $0.9\%\times\sigmag$ at the beginning of the simulations. 

In total we carry out five simulations. Their setups are shown in Table~\ref{tab:model}. The simulations have 256 grid cells in both radial (1-500~AU) and azimuthal ($2\pi$) directions. The two $1\times x\mj$ ($x$ represents the mass of planets in the model names) runs have 1 planet, the two $4\times x\mj$ runs have 4 planets, and Model~$\modelhltau$ has 3 planets. Accretion of disk material onto planets is not included in our models, and the radial locations of the planets are fixed so they don't migrate in the disk (this assumption will be discussed in Section~\ref{sec:caveats}). Planets in all models are on circular orbits. For the first 4 models, each of the neighboring pairs in the 4-planet models are locked into 2:1 resonances. These 4 models are evolved for 0.4~Myr, at which point $\sigmag$ has reached a steady state, while $\sigmabg$ at the peak of the ring outside the outermost planet changes less than $5\%$ in the final $10\%$ of time ($\sigmabg$ in models with $\geq1\mj$ planets has touched a hard floor inside the gaps in the hydro simulations). The final gas disk mass of these models is in between $0.03\msun$ and $0.04\msun$. Model~$\modelhltau$ will be discussed separately in Section~\ref{sec:hltau}.

\subsection{Monte Carlo Radiative Transfer Simulations}\label{sec:mcrt}

We carry out 3D MCRT simulations using the code developed by \citet[see also \citealt{whitney03a,whitney03b}]{whitney13}. In the simulations, the luminosity from the central star is absorbed and reemitted or scattered by the dust in the surrounding disk. The temperature in each grid cell is calculated based on the radiative equilibrium algorithm described in \citet{lucy99}. The anisotropic scattering phase function is approximated using the Henyey-Greenstein function. Polarization is calculated assuming a Rayleigh-like phase function for the linear polarization \citep{white79}. We focus on images at $H$~band and ALMA band 7 (continuum emission at $870~\micron$). SEDs and images (both full resolution and convolved images) at $10~\micron$ and $100~\micron$ are presented as well \footnote{In this work, the physical quantity recorded in all model images is the specific intensity, or intensity $I_\nu$ for short, which has the unit [mJy~arcsec$^{-2}$], or [ergs~s$^{-1}$~cm$^{-2}$~Hz$^{-1}$~arcsec$^{-2}$]. This quantity is sometimes referred to as ``spectral radiance''.}. This code has been used to model protoplanetary disks in the past (e.g. \citealt{hashimoto12,zhu12,dong12cavity,dong12pds70,follette13,grady13}).

The disk setup is largely adopted from \citet{dong12cavity}. We construct a 3D disk structure in spherical coordinates. The number of grid cells in the radial ($r$), azimuthal ($\phi$), and polar ($\theta$) directions are 399, 256, and 200, respectively. All simulations are run with 4 billion photon packages. The central source is a 1~$\msun$, 4500~K pre-main-sequence star with a surface gravity $g=10^4~{\rm m/s^2}$ and solar metallicity. The inner boundary of the disk is at the dust sublimation radius where the dust temperature reaches 1600~K ($r_{\rm sub}$, $\sim0.1~AU$), while the outer boundary is at 500~AU.


There are two disk components in our models -- a ``small'' dust particle size disk, and a ``big'' dust particle size disk. The grains in the small dust disk (``small grains'' from now on) are the standard interstellar medium (ISM) grains as in \citet{kim94}. These grains contain silicate, graphite, and amorphous carbon, and their size distributions roughly runs from $\sim0.02\micron$ to $\sim1\micron$ (a smooth power law distribution in the range of $0.02-0.25\micron$ followed by a sharp cut off beyond $0.25\micron$). Their optical properties can be found in Figure~2 in \citet{dong12cavity}. The grains in the big dust disk (``big grains'' from now on) are assumed to be 2/3 silicate (density 3.3~g/cm$^3$) and 1/3 graphite (density 2.3~g/cm$^3$). We adopt the grain properties from \citet{laor93}. The minimum and maximum grain sizes are assumed to be 0.9~mm and 1.1~mm, and the number density $n(s)$ dependence on the grain size $s$ goes as $n(s)\propto s^{-3}$ (since the size range is so narrow, the details of the size distribution have little effect on the properties of the grains). The opacity, albedo, average cosine scattering angle, and maximum polarization are calculated using the routine developed by \citet{bohren83}\footnote{We use the version of the \citet{bohren83} routine that has been modified by B. T. Draine, https://www.astro.princeton.edu/$\sim$draine/scattering.html.  These can be reproduced with a code written by T. Robitaille, https://github.com/hyperion-rt/bhmie}.

We directly read in the 2D gas and 1~mm big grain surface density from our hydro simulations to set up MCRT models. In general, the small grains are strongly coupled to the gas in our models due to their short stopping time (Eq.~\ref{eq:Ts}, $\Ts\ll1$, Figure~\ref{fig:ts_rp}). Therefore, the surface density of the small grains $\sigmasg$ is linearly proportional to $\sigmag$. Tests with $1\micron$-sized particles in our models have been carried out to confirm the validity of this assumption, and $1\micron$-sized particles are not included in standard hydro runs for computational expenses considerations. We set $\sigmasg=0.1\%~\sigmag$, so that the {\it initial} big-to-small-dust mass ratio is 9:1 (initial $\sigmabg=0.9\%~\sigmag$), and the {\it initial} gas-to-dust (including both big and small dust) mass ratio at $t=0$ in the hydro calculations is 100:1, a canonical value assumed in protoplanetary disks. Both ratios change as the distribution of 1~mm particles are evolved independently from the gas in hydro calculations, while $\sigmasg$ is always fixed as 10$^{-3}\sigmag$, though changes are small in our models as the final gas-to-dust ratio always stays within $20\%$ from the initial value (100:1). In the disk radius range that is covered by the hydro calculations, we map the 2D hydro grid onto the 2D MCRT grid with a second order interpolation scheme. For the inner disk that is not covered by the hydro models ($r_{\rm sub}\leq r<1$~AU), we extrapolate the surface density at the inner boundary of the hydro simulations inward.

We note that both the gas-to-dust mass ratio and the big-to-small-dust mass ratio are not well constrained by observations. Results have shown that the gas-to-dust mass ratio can be quite low (as low as $\sim$20) in some Class II disks \citep{williams14}. For the big-to-small-dust mass ratio, any number from $\sim0$ (the ISM value) to $\infty$ (solids completed converted to planetesimals) may exist, and our choices of 9:1 just represents a non-special middle stage. We note that we do not include dust evolution processes such as coagulation or fragmentation. They are crucial in obtaining a long-term self-consistent grain size distribution (e.g. \citealt{birnstiel10}), and are beyond the scope of this study. The NIR and mm images in our models are more or less determined by the distribution of small and big grains independently, easing the analysis of the model results. As a consequence of the uncertainties of the two ratios, absolute values of $\sigmasg$ and $\sigmabg$ and the resulting absolute intensities of the images should be taken as references only. The absolute flux at NIR is less affected by these choices, as changing $\sigmasg$ by a few orders of magnitudes in systems like ours may only change the NIR flux by a factor of a few (see Figure~5 in \citealt{dong12pds70} for one example, also see \citealt{dong12cavity}), while optically thin mm continuum flux is roughly proportional to $\sigmabg$ within a reasonable range. Relative intensities (i.e. contrast of the gaps in images) at a given wavelength are more robust as they are less affected by the choices of these ratios.


Once we have the 2D distributions for both grain-model disks, we need to vertically extend the disk to construct 3D structures for the MCRT simulations. This is done by assuming Gaussian profiles for the volume density of both grains in the vertical direction,
\begin{align}
   \rho_{\rm sg}(z)&=\frac{\sigmasg}{\hsg\sqrt{2\pi}} e^{-z^2/2\hsg^2}, \label{eq:rhosg}
\\ \rho_{\rm bg}(z)&=\frac{\sigmabg}{\hbg\sqrt{2\pi}} e^{-z^2/2\hbg^2}, \label{eq:rhobg}
\end{align}
where $\hsg$ and $\hbg$ are the scale height of the small and big grains, respectively. As small grains are well mixed with the gas, $\hsg=\hg$. On the other hand, big grains tend to settle toward the disk mid-plane, and their vertical distribution is determined by the balance between gravitational settling and turbulent diffusion. As in \citep{cuzzi93} and \citep{youdin07},
\begin{equation}
\hbg=\frac{\hg}{\sqrt{1+\Ts Sc/\alpha}}
\label{eq:hbg}
\end{equation}
where the dimensionless stopping time $\Ts$ is
\begin{equation}
\Ts=\ts\Omega.
\label{eq:Ts}
\end{equation}
We set $\alpha=0.001$ and $Sc=1$ as in the hydro models (Section~\ref{sec:hydro}). We caution that in MRI turbulent disks
dominated by ambipolar diffusion, Sc can be larger than 1 \citep{zhu14dust}.

The full-resolution model images produced by the MCRT simulations need to be convolved by a PSF in order to achieve an angular resolution comparable to observations. We convolve the $H$~band images using a circular Gaussian kernel with a full width half max (FWHM) of $0.04\arcsec$ (6~AU at 140~pc), as a good approximation to the angular resolution achieved by Subaru, VLT, and Gemini with their high contrast imaging systems (the FWHM of an airy disk is $1.028\lambda/D\sim0.04\arcsec$ for a primary mirror with a diameter $D=8.2$~m at $\lambda=1.6~\micron$). For the first 4 models, convolved images at ALMA band 7 are produced by convolving the raw images by a Gaussian kernel with a FWHM of $0.1\arcsec$ (i.e. a $0.1\arcsec\times0.1\arcsec$ beam, 14~AU at 140~pc), a typical beam size routinely achievable by ALMA in the near future. The raw ALMA image for Model~$\modelhltau$ is convolved by a Gaussian kernel with a FWHM of $0.035\arcsec$ for finer resolution (see Section~\ref{sec:hltau}). The source is assumed to be at a distance of 140~pc from earth. We note that realistic instrumental effects such as flux loss and observational noise, which may affect both the absolute and relative flux, are not included in the production of our model images (beyond the intrinsic Monte Carlo noise in the radiative transfer images).


\section{Results}\label{sec:results}

In this section, we present results from the hydro and MCRT simulations for Models~$\modelzero$, $\modelone$, $\modeltwo$, and $\modelthree$ (Model~$\modelhltau$ will be discussed separately in Section~\ref{sec:hltau}), including surface density maps for both grain models, and images at two inclinations for $H$~band and ALMA band 7 (continuum at 870~$\micron$). The SEDs and raw images at $10~\micron$ and $100~\micron$ are shown in Appendix~\ref{sec:appendix}. Throughout the paper, we use a ``blue-hot'' color scheme for small grains and scattered light related presentations, and ``red-hot'' color scheme for big grains and dust thermal emission related presentations.

\subsection{Density Structure of the Models}\label{sec:results-density}

The 2D surface density distributions for both the big and small grains are shown in Figure~\ref{fig:sigma}, and their azimuthally averaged radial profiles are shown in Figure~\ref{fig:sigma_rp}. Gap property measurements are listed in Table~\ref{tab:sigma_measurement}.

The gap density contrast in the small and big grains differ dramatically in all cases. Defined as the ratio of the peak azimuthally-averaged surface density in the outer disk, $\Sigma_{\rm max,out}$, to the minimum value inside the gap, $\Sigma_{\rm min,gap}$, the gap density contrasts $\zeta_\Sigma$ are listed in Table~\ref{tab:sigma_measurement}. In Models~$\modelone,\ \modeltwo$ and $\modelthree$, $\zeta_\Sigma\gtrsim10^9$ for the big grains\footnote{Note that  $\sigmabg$ has a hard floor in the hydro simulations in order to keep the runs stable, so $\zeta_\Sigma$ for the big grains could be higher.} due to a strong dust filtration effect (the dimensionless stopping time $\Ts$ is around 0.1 for 1~mm particles at the outer gas gap edges) while it is only $\sim$1.5 orders of magnitude for the small grains. In Model~$\modelzero$, where we have a 0.2$\mj$ planet, the gap is almost flat in the small grains, while still more then 3 orders of magnitude deep in the big grains.

The width and position of the gap depends on the configuration of the planetary system. In the small grains, a narrow or almost no gap is opened up in the single-planet models, as small grains are allowed to cross the orbit of the planet and populate the inner disk.  In contrast, gaps opened by each of the individual planets in the four-planets models overlap with each other and form a wide common gap.  For the big grains, the gas pressure bump outside the outermost planet's orbit in all 4 models effectively traps the grains and piles them up. Increasing the mass of the planets in the $4\times x\mj$models from $1M_{\rm J}$ to $2M_{\rm J}$ causes the gas gap to become wider and the edge becomes sharper, so the ``ring'' in the big grains becomes wider and moves outward. Also, in Model~$\modelthree$, the gap is slightly eccentric, as a result of the disk-planet interaction in the high planet mass regime \citep{kley06, dangelo06, ataiee13, pinilla14}. In all three cases with $\mplanet\geq\mj$, density waves from the planets and streamers inside the gap are clearly visible in $\sigmasg$, while they are only marginally traceable in the distribution of big grains in some cases (e.g. Model~$\modeltwo$).

To illustrate the dust settling effect, the vertical density structure of both grains in Model~$\modelone$ at $\phi=0^\circ$ is shown in Figure~\ref{fig:rho_rz}. While the small dust disk has a flared structure, the big grains collapse to the mid-plane, especially at regions where $\sigmag$ is low. Within the gap, the gas surface density drops significantly, leading to a even larger $T_{s}$ and smaller $\hbg$.


\subsection{Face-on Disk Images}\label{sec:results-faceon}

Figure~\ref{fig:image_theta0} shows the face-on MCRT model images (i.e. at a viewing angle $\theta=0^\circ$) at both $H$~band (polarized intensity) and ALMA band 7 (continuum emission at 870~$\micron$). The azimuthally averaged radial profile of these images are shown in Figure~\ref{fig:image_theta0_rp}\footnote{Model images are binned into annuli with a width of $\sim$2~AU.}, and measurements of the gap properties are listed in Table~\ref{tab:image_measurement}.

The gap size at the two wavelengths are very different. The mm images have a similar structure in all models: a bright ring, a centralized peak, and a gap in between. The emission signals from the central region are mostly due to the small grains in the $4\times x\mj$ models and the big grains in the $1\times x\mj$ models (the emission from the central peak is unreal as it depends on how the inner $\sim1$AU is filled up in the MCRT pre-processing step). As expected, the mm dust continuum emission in the outer disk closely follows the distribution pattern of the big grains (note the slightly eccentric ring in Model~$\modelthree$). Inside the gap, $\sigmabg$ is too low in models with $\mplanet\geq1\mj$, and the floor mm emission intensity is set by the small grains. In contrast, the scattered light images of Model~$\modelzero$ show almost no gap, while $\modelone$ shows a narrow gap, and the two $4\times x\mj$ models reveal a much wider gap. Measurements of the gap sizes in the convolved images are listed in Table~\ref{tab:image_measurement}, where the outer gap edge ($r_{\rm gap, out, image}$) and inner gap edge ($r_{\rm gap, in, image}$) are defined to be the locations in the gap where the intensity reaches half of its peak value in the outer disk ($I_{\rm max, out}$). With $1~M_{\rm J}$ planets in both cases, $\deltagap$ ($r_{\rm gap, out, image}-r_{\rm gap, in, image}$) at $H$ band in Model~$\modelone$ is only $45\%$ of $\deltagap$ in Model~$\modeltwo$.

Gaps in scattered light are much shallower than in thermal emission. This is caused by (1) the smaller gap contrast in $\sigmasg$ then in $\sigmabg$, and (2) the fact that scattered light is only probing disk surface features \citep{dong12pds70, dong12cavity}. Similar to $\zeta_\Sigma$, we define ``gap contrast'' in raw and convolved images, $\zeta_{\rm image,raw}$ and $\zeta_{\rm image,conv}$, as the ratio of the maximum azimuthally averaged intensity in the outer disk ($I_{\rm max, out}$) to the minimum value inside the gap ($I_{\rm min, gap}$), and list the measurements in Table~\ref{tab:image_measurement}. At $H$~band, $\zeta_{\rm image,conv}$ are only around 3-6 for all models with $\mplanet\geq1\mj$, comparing with $\sim100-1600$ at mm wavelengths. The gap contrast difference in the raw images are even bigger. 

Another feature at $H$~band is the visibility of density waves and streamers. They are very clear in raw images, and marginally traceable in convolved images with much less contrast (particularly in Model~$\modelthree$, bottom row in Figure~\ref{fig:image_theta0}; note that observational noises, which are not added in our convolved images, may further weaken the strength of the density waves). On the other hand, spiral density wave are not evident at mm wavelengths. Spiral arm like features have been found in scattered light images in recent years \citep[e.g.][]{muto12, grady13,benisty15}, and yet their origins are still largely unknown \citep{juhasz14}. We defer a detailed study of the appearance of the spiral arms to the next paper.


Lastly, the location of the outer gap edges in the model images depend on wavelength, as pointed out by \citet{dejuanovelar13} and \citet{pinilla14}. The radii of the peak intensities in the outer disk in the azimuthally averaged convolved images are indicated in Figure~\ref{fig:image_theta0_rp} (listed in Table~\ref{tab:image_measurement} as $r_{\rm max, out, image}$). In all cases, the peak in the mm is at a larger radius than at $H$~band. In addition, the difference between them increases as the mass of the planets increase from Model~$\modeltwo$ to Model~$\modelthree$. Note that this is not due to the difference in the locations of the peak surface density in the outer disk in the two grains, as $r_{\rm max, out, \Sigma}$ is actually slightly smaller in $\sigmabg$ than in $\sigmasg$ (Figure~\ref{fig:image_theta0} and Table~\ref{tab:sigma_measurement}). This difference is mainly caused by a radiative transfer effect. As the mm emission linearly scales with $\sigmabg$, $r_{\rm max, out, image}$ at mm wavelengths closely traces $r_{\rm max, out, \Sigma}$ in the big grains. On the other hand, the NIR scattered light more closely traces the {\it abrupt changes} in $\sigmasg$, as they lead to sudden variations in the shape (curvature) of the disk surface. As a result, the $H$~band images peak around the outer gap edge in $\sigmasg$, not $r_{\rm max, out, \Sigma}$ in the small grains.


\subsection{Inclined Disk Images}\label{sec:results-inclined}

Figure~\ref{fig:image_theta45} shows model images at a viewing angle $\theta=45^\circ$. The $H$-band images show a large scale asymmetry along the minor axis: the top (far) side of the disk appears to be fainter than the bottom (near) side of the disk. This is due to forward scattering from the dust particles in an inclined disk (\citealt{henyey41}, face on images and inclined mm images do not have this asymmetry). In addition, the bright elliptical ring is off center from the star. This is due to shadowing and geometric effects (see also \citealt{thalmann14}). 

To illustrate these effects, we run a control model that has the same density structure with Model~$\modeltwo$ at $r\geq35$~AU, but no material inside 35~AU ($\modeltwo$-empty-gap), and compare $H$~band images of the two in Figure~\ref{fig:image_emptygap}. The raw image of Model~$\modeltwo$ (the upper left panel) reveals a dark lane in the middle on the far (top) side of the gap wall (indicated by the arrow). This is because the material inside the gap, including the residual inner disk, spiral arms, and streamers, block the star light from reaching the middle part of the gap wall \citep{espaillat11}. Due to the low surface density in the gap region and the flaring of the disk, the inner disk does not block the starlight from reaching the upper and lower part of the wall. Consequently, when convolved, the bright ring roughly overlaps with the upper edge of the gap wall, which is not at the same plane as the star (upper right panel). Therefore, if fitting the ring by an ellipse (the dashed red ellipse), the center of the ellipse offsets from the star (marked as $\times$). In our example here, the offset ($\sim5$AU) is $\sim8\%$ the length of the minor axis, or $\sim12\%$ of the gap radius. On the other hand, in Model~$\modeltwo$-empty-gap, the entire outer gap wall is uniformly illuminated (i.e. no dark middle lane on the far side, lower left panel), and the center of a fitted ellipse to the ring in the convolved image almost coincides with the star (lower right panel; a small offset $\sim0.01\arcsec$ along the minor axis still exists due to a geometric effect, as only the upper edge of the near side gap wall is visible while the entire far side gap wall is visible).


\section{Discussion}\label{sec:discussion}

\subsection{Comparison between Model Results and Observations}\label{sec:comparison}

To date, a number of transitional disks have been resolved at NIR and/or mm wavelengths. Here we compare the morphology and properties of the observed systems with our models (Model~$\modelhltau$ will be discussed separated in Section~\ref{sec:hltau}), aiming at answering the basic question of whether gaps opened by single or multiple planets are broadly consistent with observations or not.

\subsubsection{Gap size}\label{sec:gapsize}

The giant gap/cavity revealed in NIR imaging of a few transitional disks (e.g. RX~J1604.3-2130, \citealt{mayama12}; HD~142527, \citealt{canovas13}; SAO~206462, \citealt{garufi13}; and PDS~70, \citealt{hashimoto12}) are all quite large, ranging from $\sim28$~AU in SAO~206462 to over 100~AU in HD~142527. As the width of the NIR gap opened by a $1\mj$ planet at 30~AU in our model is only 10~AU, the single-planet scenario may face major difficulties in explaining observations\footnote{It might be possible that a single massive companion with a mass close to or exceeding the brown dwarf mass limit of 13~$\mj$ could open a big enough gap, see \citet{dejuanovelar13} and \citet{artymowicz94}.}. Note that having a bigger planet is not a good solution as $\deltagap$ only depends on $\mplanet$ weakly. As \citet{fung14} pointed out, $\Delta_{\rm gap,\sigmag}\sim2\times$~max$(R_{\rm Hill}, \hg)$, where $R_{\rm Hill}\propto r_{\rm p}\mplanet^{1/3}$ is the Hill radius of the planet (see also Figure 5 in \citealt{duffell13}). The observed gap sizes are more consistent with our four-planets models (Table~\ref{tab:image_measurement}). We note that a single body with a mass around the brown dwarf lower mass limit ($\sim$13~$\mj$) may be able to create a big gap in NIR images, as shown in \citet{dejuanovelar13}. Also, the size of the gap depends on the location of the planets ($R_{\rm Hill}\propto r_{\rm p}$ and $\hg\propto r^{1+\delta}$, where $\delta\sim0.25$), so shifting planets outward will increase the gap sizes.

\subsubsection{Gap Depth}\label{sec:gapdepth}

The gap depth in observations is determined by the configuration of the planets, the properties of the disk, radiative transfer processes, and image convolution in a complicated way. In total there are 4 transitional disks whose giant gap has been imaged at NIR wavelengths with excellent image quality to facilitate the same gap contrast measurement as we performed to our models. We compare their observed values with the two 4-planet models in Table~\ref{tab:nirgapcontrast}. The quoted values and errors are the average and standard deviation over different azimuthal angles. We note that in three out of four observed systems (except HD~142527) the possible inner edge of the gap (if there is one) is not detected as it is blocked by the inner working angle; therefore the measured value for these systems should be considered as a lower limit. On the other hand, observational noise, which may reduce the contrast of the images, is not added in our model images, so the measured values may be upper limits (Monte Carlo images also have noise from limited photon numbers, but it may be lower than the observations). Nevertheless, our models show promising agreement with observations. 

In the mm/mm interferometric observations, the gap contrast has to be constrained through model fitting of the disk visibility. In the past, the thermal emission inside the gap was often not resolved and in most cases was below the noise floor set by the instrument sensitivity. In these cases, mm data can provide only weak upper limits on the gap contrast in either the emission intensity or $\sigmabg$. This situation is changing as ALMA is starting to detect and spatially resolve the ``residual'' dust emission inside the gap as it is transitioning into final phase. Recently, \citet{vandermarel15} modeled both continuum and gas line emissions in 6 transitional disks with ALMA data. The drop in dust surface density inside the cavities ($\delta_{\rm dustcav}$ in their definition) was concluded to be at least 10$^{-4}$ for most objects, and even as low as $10^{-6}$ in the case of RX~J1604.3-2130. This is broadly consistent with our hydro results (an $\sim$8 order of magnitude drop in $\sigmabg$ in the multi-planet cases). In reality, dust particles with sizes from sub-$\micron$ to pebbles will co-exist in the disk, and all contribute to the mm opacity. We note that inside the deep gaps in our models small grains actually contribute most of the mm optical depth (i.e. $\sigmasg\kappa_{\rm sg}>\sigmabg\kappa_{\rm bg}$, where $\kappa$ denotes dust opacity) despite their low opacity comparing with the big grains, as the big grains are so heavily depleted from the gap. Therefore the drop in mm ``opacity'' inside the gap, as often the quantity constrained from mm data modeling, can be smaller than the drop in $\sigmabg$. Detailed modeling with a size distribution from realistic dust evolution models are needed to make more insightful comparisons between models and specific systems.


\subsubsection{Gap Size Dependence on Wavelengths}\label{sec:gapdepth}

One class of transitional disks have a clear cavity at mm, but no gap/cavity down to the inner working angle in NIR imaging (\citealt{dong12cavity}, e.g. SR~21, \citealt{perez14, follette13}). For these objects, our models suggest two possible explanations involving gaps opened by planets: (1) the NIR gap may be too shallow and/or too narrow to be detectable due to small planet masses (i.e. Model~$\modelzero$, which has almost no detectable gap at NIR and a mm gap with a contrast of $\sim10$). (2) The NIR gap may be entirely hidden under the inner working angle. These solutions were first proposed by \citet{pinilla12b} and \citet{dejuanovelar13} with in depth 2D gas dynamics simulations and 1D dust evolution models. We confirm the results of these previous works with our calculations.

RX~J1604.3-2130 \citep{zhang14} and SAO~206462 \citep{garufi13} both show a bigger cavity size at mm than at NIR ($\sim78\ vs\ \sim63$~AU in RX~J1604.3-2130, and $\sim46\ vs\ \sim28$ in SAO~206462), while HD~142527 shows a bigger cavity size in big grains than in the gas (\citealt{perez14hd142527}, $\sim140$~AU vs $\sim90$~AU). This is consistent with the bigger cavity sizes seen at mm than at NIR in our model images (Table~\ref{tab:image_measurement}). For example, in Model~$\modelthree$ the distance between the outermost planet at 30~AU and the outer gap edge is two times higher in the convolved mm image (15~AU) than in the convolved NIR image (8~AU). 


\subsection{An HL-Tau-Like System with Multiple Separate Gaps and Rings}\label{sec:hltau}

Inspired by the recent release of the ALMA image of the HL Tau disk \citep{brogan15}, we have included an additional disk-planet model, shown as Model~$\modelhltau$ in Table~\ref{tab:model}. This model has 3 planets, located on circular orbits at 12, 30, and 65 AU from the star. The initial gas surface density $\sigmag$ is 5 times higher than the profile assumed for other models (Equation~\ref{eq:sigmags}), and the system is evolved for a shorter period, 0.2~Myr, suggested by the relatively high mass (\citealt{kwon11}) and young age of the system (although we only run the hydro simulation to 0.2 Myr, the dust distribution in this model has already reached a steady state). The total gas disk mass is 0.17 $\msun$ at the end of the hydro simulation, and the final gas-to-dust mass ratio is 90:1. The synthetic mm image is convolved by a $0.035\arcsec\times0.035\arcsec$ Gaussian beam, to match the angular resolution of ALMA observations. All the other conditions in the hydro and MCRT simulations are the same as in Section~\ref{sec:setup}. The surface density maps and the synthetic images at both $H$~band and ALMA band 7 are shown in Figure~\ref{fig:hltau2d}, while the radial profile measurements are shown in Figure~\ref{fig:hltau1d}. A comparison between the model mm image and the observation of HL Tau is shown in Figure~\ref{fig:comparison}.

In the gas (small dust) disk, the inner two planets each open a shallow, narrow gap around their orbits. The perturbation induced by the outermost planet is only marginal, due to the large dynamical (gap opening) time scale at its large distance. Somewhat deeper but still narrow gaps are opened in the disk of the big grains. The outer-disk-to-gap density contrast ratio $\zeta_\Sigma$ for big grains is 28 for the innermost gap, 10 for the middle gap, and 2.5 for the outermost gap, much smaller than the other models. This is because the big grains are only marginally coupled to the gas in this case, mostly due to the high gas surface density and the shallowness of the gas gaps opened by low mass planets. The dimensionless grain stopping time $T_{\rm s}$ is only on the order of $\sim0.01$ to 0.001 around the three gap regions. As a result, the depletion of big grains inside the gap is very incomplete, and the piling up of big grains around the gas pressure peak is insignificant. In addition, in contrast to the two $4\times x\mj$ models, the gaps do not overlap with each other, and they are well separated by more or less unperturbed disk rings. This is mainly due to the large separations between the planets, and the above mentioned weak coupling between the big grains and gas.

The raw mm images of the system at both face-on and $45^\circ$ inclination angles clearly show 3 narrow gaps, separated by 2 bright rings, and an inner and an outer disk, closely matching the surface density pattern in the big grains as expected. The beam size, $0.035\arcsec\times0.035\arcsec$ (or $\sim5$~AU), is small enough, so that the convolved mm images successfully preserve these features. In our specific model, the gap contrasts $\zeta_{\rm image,conv}$ in the convolved mm image at face-on angle are 1.5, 3.8, and 1.8 from the innermost to the outermost gaps. We note that $\zeta_{\rm image,conv}$ sensitively depends on the dust-gas coupling effect, which is set by the surface density of the gas in the disk, and also the profiles of the gas gaps opened by planets, which are determined by the mass of the planets, as well as the viscosity and scale height of the disk. 

At $H$~band, similar to Model~$\modelzero$, the gaps are clear in the raw images, but are somewhat smeared out and are only marginally visible in the convolved images. Density waves excited by these low mass planets are marginally visible in the raw images, but almost invisible in the convolved images, and are not in the mm images.

\subsection{Caveats}\label{sec:caveats}

In this section we discuss additional factors that could potentially affect the gap profile in the hydro simulations, and which are not included in our hydrodynamical models.

Firstly, the interactions between a planet and the disk could drive the planet to migrate. When protoplanets open gaps in the disk, planets tend to migrate on a time scale set by the disk's viscosity (type II migration, \citealt{ward97, hasegawa13}). The migration time scale for planets in our models is typically on the order of $\sim$Myr or longer, much longer than the gap opening time scale.  Therefore, it is safe to ignore the planet migration in the study of the observational signatures of these gaps. Also, when multiple giant planets migrate together in a gaseous disk, they tend to lock each other into 2:1 mean motion resonance \citep{pieren08, zhu11}.  As a result the system tends to be stable from planet-planet interactions.

Secondly, the choice of disk scale height and viscosity will affect the gap depth \citep{fung14}. Also, magnetorotational instability (MRI) has been suggested to take place in some part of a protoplanetary disk and serve as the source of disk viscosity as well as dust diffusion. Comparing with hydro simulations that have the same nominal $\alpha$, gaps opened by planets in MRI simulations tend to be deeper and wider \citep{nelson03, zhu13}. We defer detailed studies of gap morphology dependence on these factors to future studies.

Thirdly, accretion of disk material onto planets may happen as planets grow. Allowing planets to accrete from the disk will make a difference on the gap depth, as the material inside the gap may be drained onto the planets in addition to be cleared due to tidal forces. The more efficient accretion is, the cleaner the gap is \citep{zhu11}. Planetary accretion may also introduce distortion in the distribution of gas and dust close to the planets \citep{owen14}. Since the efficiency of planetary accretion is still largely not well understood, we choose not to take it into account and to focus on the non-accreting cases.

Last, we note that our choices of the gas-to-dust mass ratio, the big-to-small-dust mass ratio, and the employment of a simple Gaussian PSF kernel without including observational noises may affect the MCRT processes and subsequently the flux in the final images, as discussed in Section~\ref{sec:mcrt}. Also, we do not include dust evolution processes such as coagulation or fragmentation, as in \citet{pinilla12b} and \citet{dejuanovelar13}), as we expect that it is not likely to have a major impact in the results here due to the short evolutionary timescale of the models.



\section{Summary}\label{sec:summary}

Theoretical studies in the past have shown that multiple planets are able to create wide gaps in the gas surface density \citep{zhu11, dodsonrobinson11}, which resemble the appearance of cavities in transitional disks \citep{espaillat14}. The basic questions of whether these density gaps can be seen in observations at various wavelengths, and if they are broadly consistent with observed disk properties, have only been partially addressed \citep[e.g.][]{pinilla12b,dejuanovelar13,pinilla14,gonzalez12}. By combining 2D two fluid gas + particle hydrodynamic calculations with fully 3D Monte Carlo Radiative Transfer simulations, we explore further a number of observational signatures of gaps opened by one or more planets in protoplanetary disks. We produce images at $H$~band and mm wavelengths with realistic angular resolutions, and compare them with resolved observations of transitional disks.

\begin{figure}
\begin{center}
\includegraphics[trim=0 0 0 0, clip,scale=0.8,angle=0]{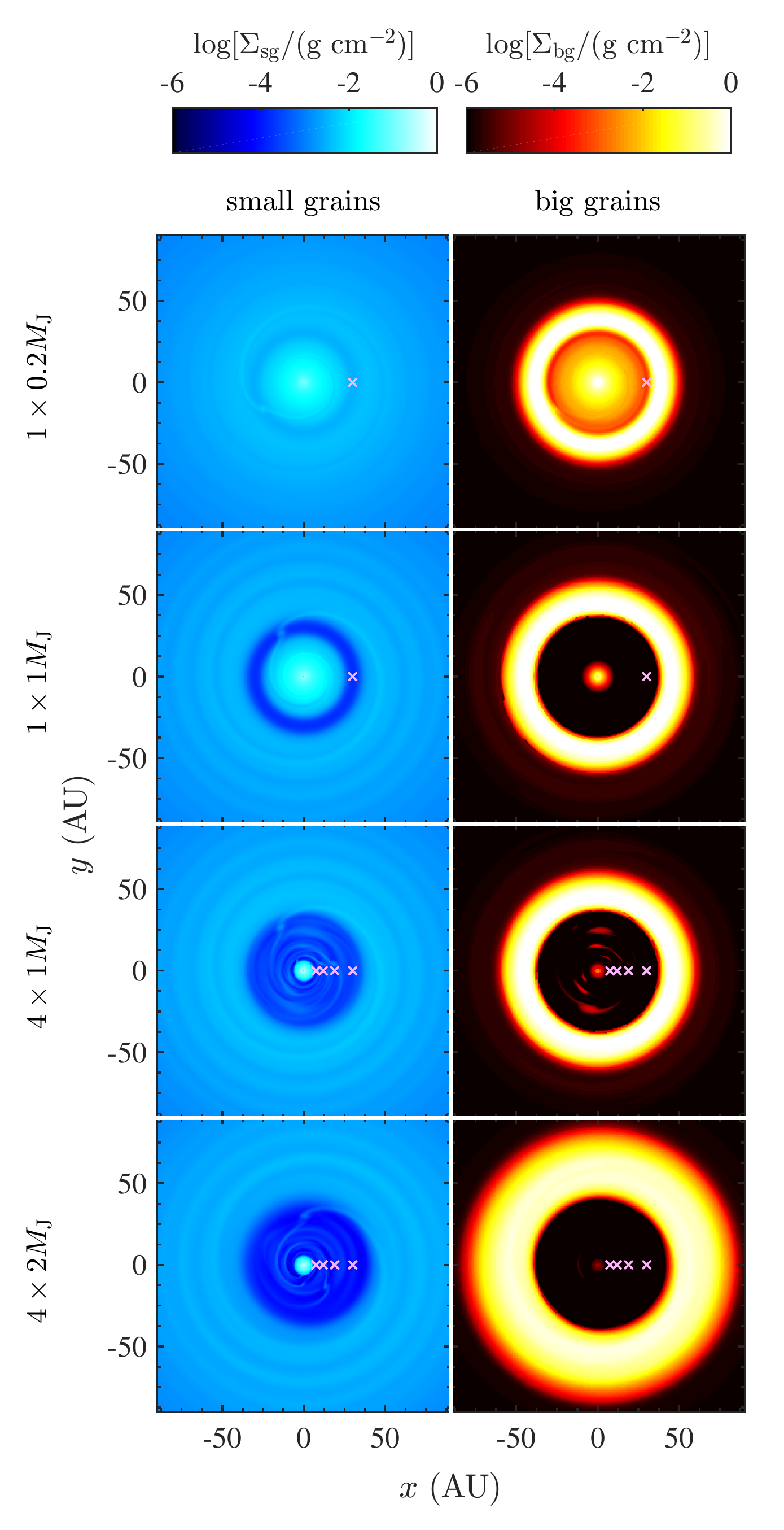}
\end{center}
\figcaption{2D surface density maps for the small (left column) and big grains (right column) from the hydro simulations (model names labeled on the left). The grey crosses in the convolved images mark the orbits of the planets. The left column is also the scaled surface density distribution of the gas, as we assume that the small grains are well mixed with the gas, $\sigmasg=0.1\%\times\sigmag$.
\label{fig:sigma}}
\end{figure}

\begin{figure}
\begin{center}
\includegraphics[trim=0 0 0 0, clip,scale=0.7,angle=0]{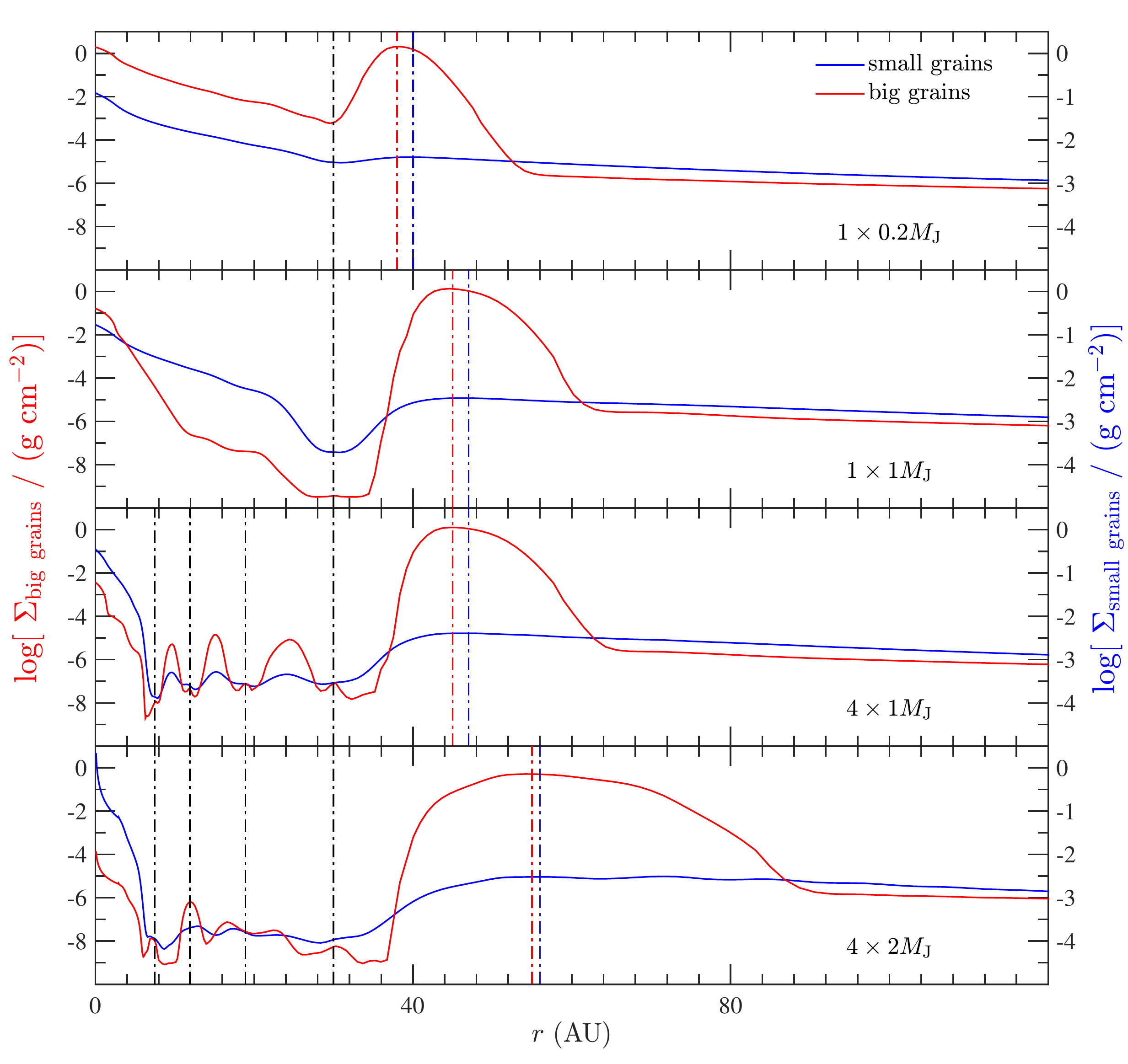}
\end{center}
\figcaption{The azimuthally averaged surface density radial profile for both the small and big grains from the hydro models (model names labeled on the left). Note that the $y$-axis tick mark labels on the left are for the big grains and the ones on the right are for the small grains. The vertical black dash-dot lines indicate the positions of the planets in each model. The vertical blue and red dash-dot lines mark the positions of the peaks in $\sigmasg$ and $\sigmabg$ in the outer disk, respectively ($r_{\rm max, out, \Sigma}$ in Table~\ref{tab:image_measurement}). The gap contrast (defined as the ratio of the peak surface density in the outer disk to the floor value inside the gap, $\zeta_\Sigma$ in Table~\ref{tab:sigma_measurement}) is about 1.5 orders of magnitude for the small grains in cases with $\mplanet\geq1~\mj$, and $\sim9$ orders of magnitude for the big grains. Note that $\sigmabg$ has a hard floor in the hydro simulations, so the gap contrasts for the big grains are lower limits.
\label{fig:sigma_rp}}
\end{figure}

\begin{figure}
\begin{center}
\includegraphics[trim=0 0 0 0, clip,scale=0.7,angle=0]{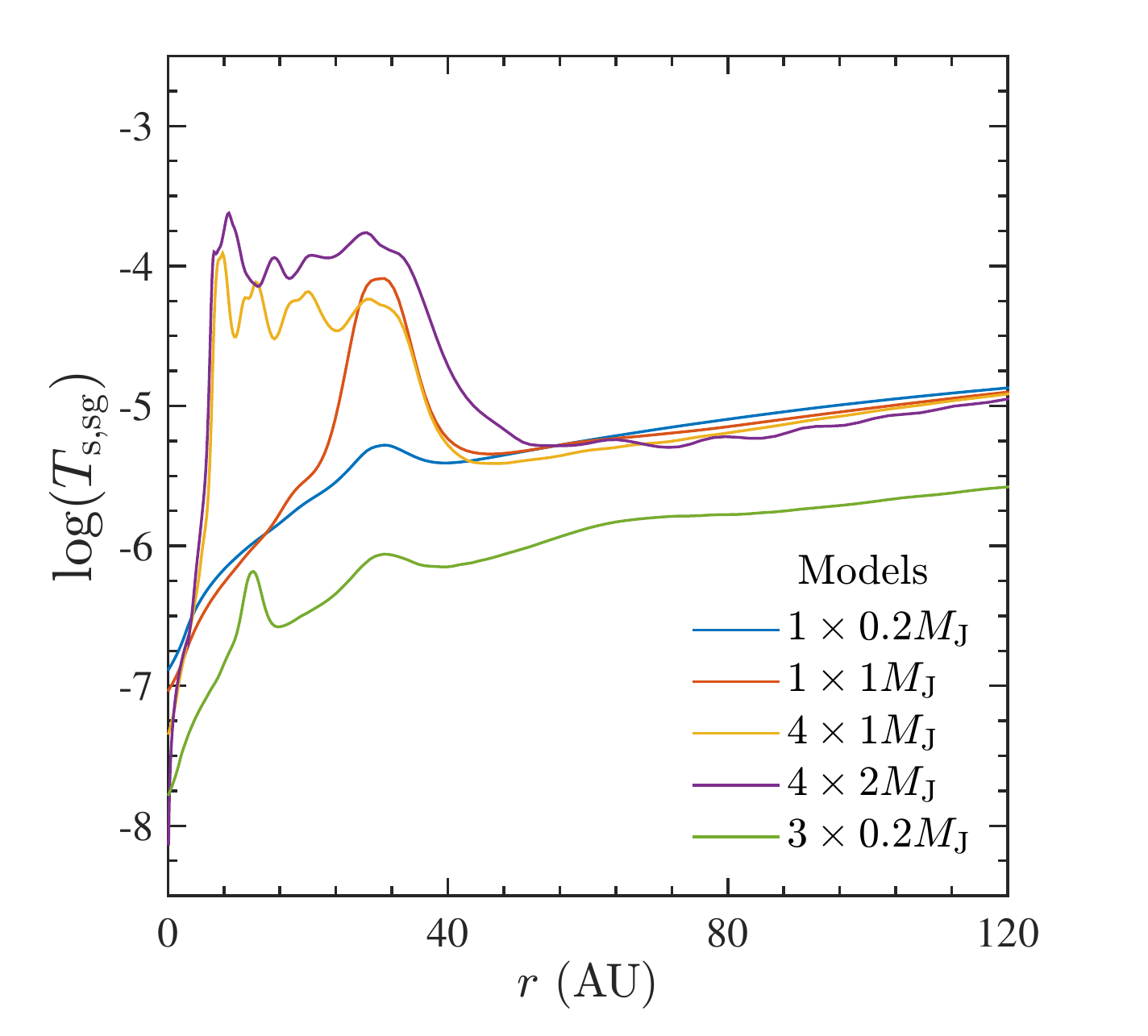}
\end{center}
\figcaption{The dimensionless stopping time $\Ts$ (Eq.~\ref{eq:Ts}) for the small grains (grain size $s$=0.1~$\micron$ as a representative size of the size distribution, Section~\ref{sec:mcrt}). For all models $T_{\rm s,sg}\ll1$.
\label{fig:ts_rp}}
\end{figure}

\begin{figure}
\begin{center}
\includegraphics[trim=0 0 0 0, clip,scale=0.75,angle=0]{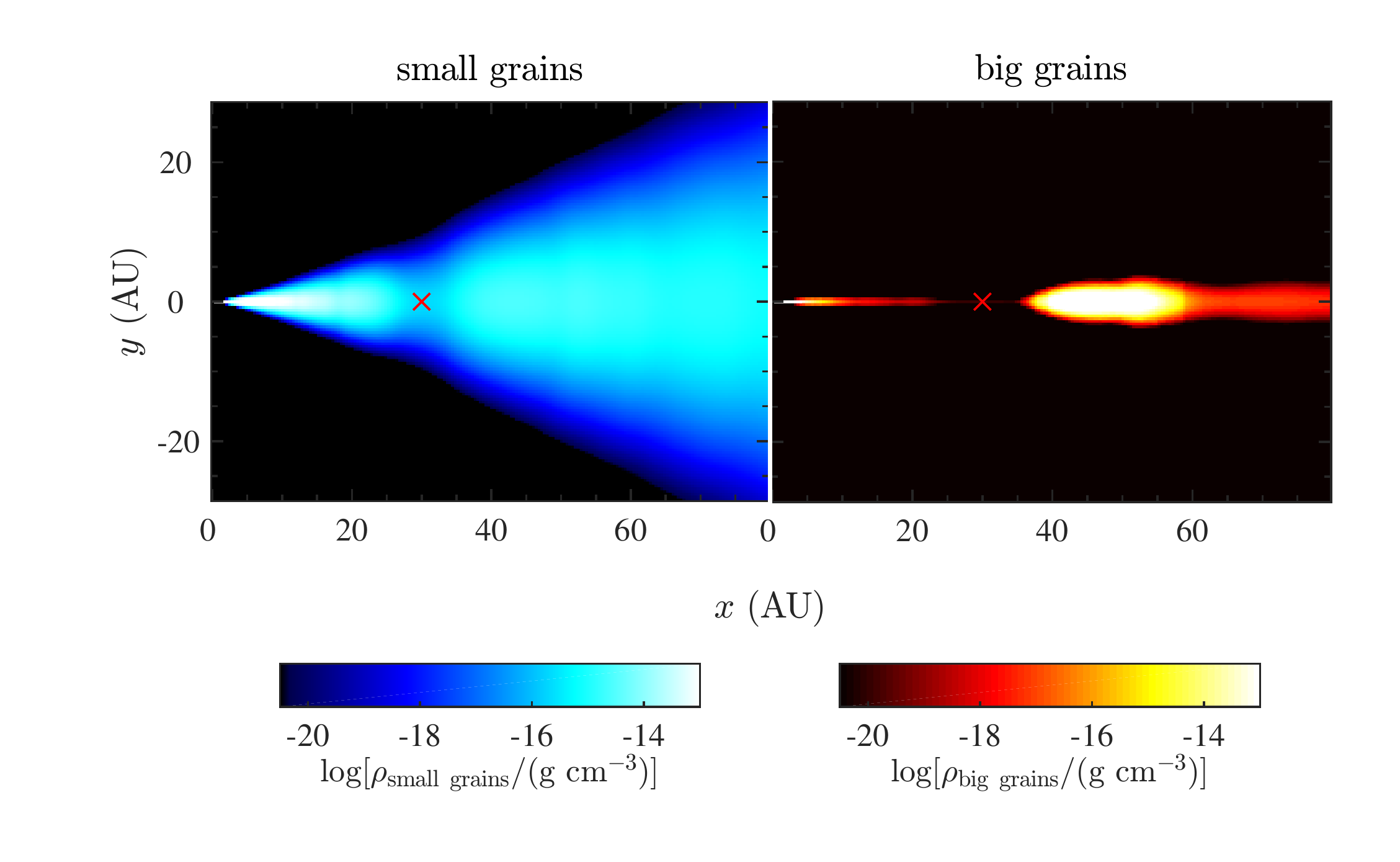}
\end{center}
\figcaption{2D volume density map for the small (left) and big (right) grains showing the vertical structure ($r-\theta$ plane) at an azimuthal angle $\phi=0^\circ$ in Model $\modelone$. The red cross indicates the position of the planet in each panel. The small dust disk is extended in the vertical direction and has a flared structure, while the big grains settle to the disk mid-plane due to aerodynamics effects. The gap around 30 AU is clearly visible in both grains.
\label{fig:rho_rz}}
\end{figure}

\begin{figure}
\begin{center}
\includegraphics[trim=0 0 0 0, clip,scale=0.75,angle=0]{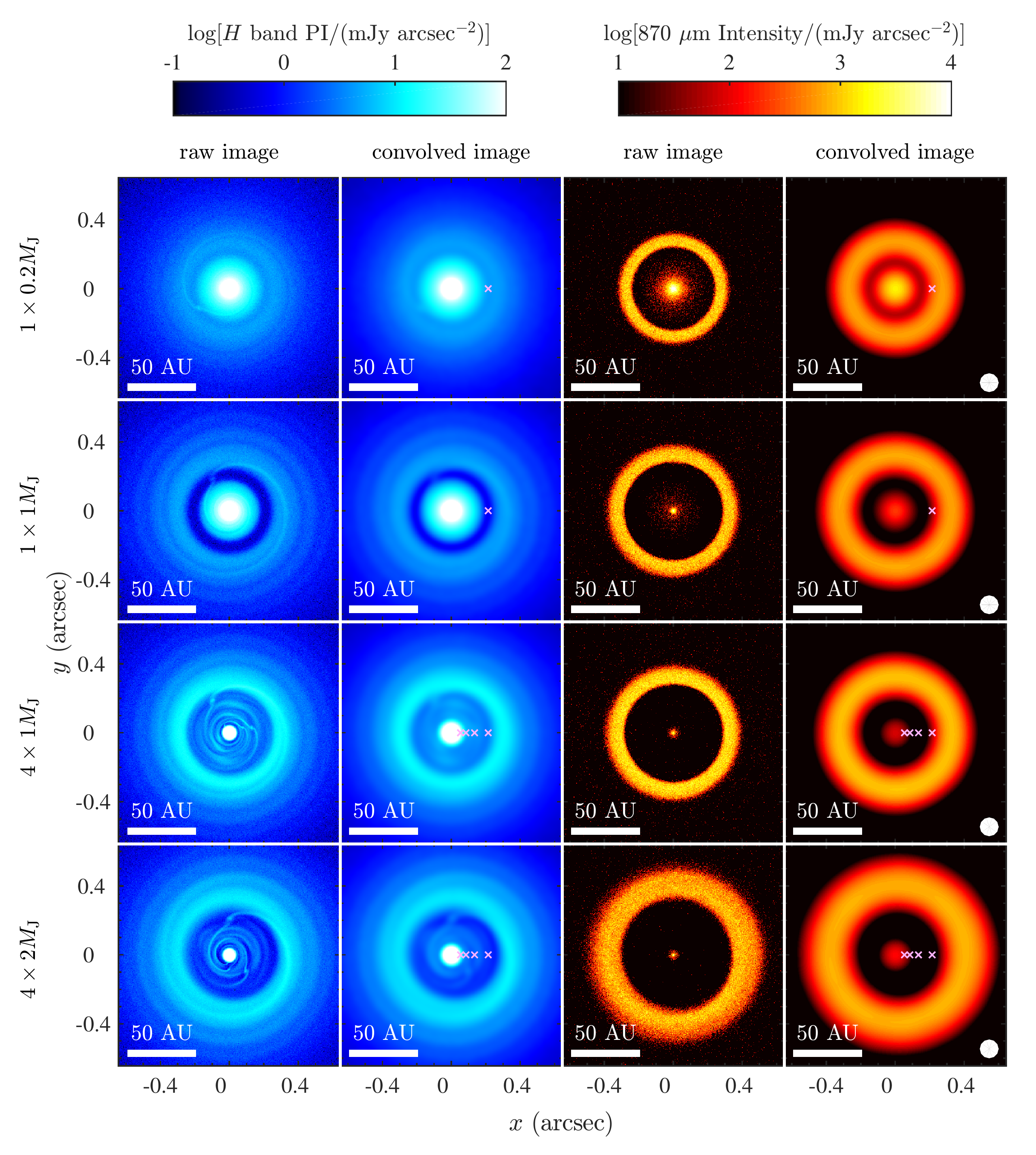}
\end{center}
\figcaption{Raw and convolved $H$~band polarized intensity images (left two columns) and ALMA band 7 (870$~\micron$ continuum) intensity images (right two columns) for our models (model names labeled on the left) at face on angle $\theta=0^\circ$. Systems are assumed to be at 140~pc. The MCRT images are convolved by a Gaussian kernel with FWHM=0.04$\arcsec$ at $H$~band and FWHM=0.1$\arcsec$ at ALMA band 7 (beam size indicated at the lower right corner in the convolved mm images) to mimic realistic angular resolutions. The grey crosses in the convolved images mark the orbits of the planets. While models with four $\mplanet\geq1\mj$ planets (the bottom two rows) all have a wide gap at both NIR and mm, in Model~$\modelone$ (the second row) the NIR gap is much smaller than the mm gap. The NIR gap in Model~$\modelzero$ (the top row) may be to weak to be detectable, while the gap at mm wavelengths is significant. 
\label{fig:image_theta0}}
\end{figure}

\begin{figure}
\begin{center}
\includegraphics[trim=0 0 0 0, clip,scale=0.75,angle=0]{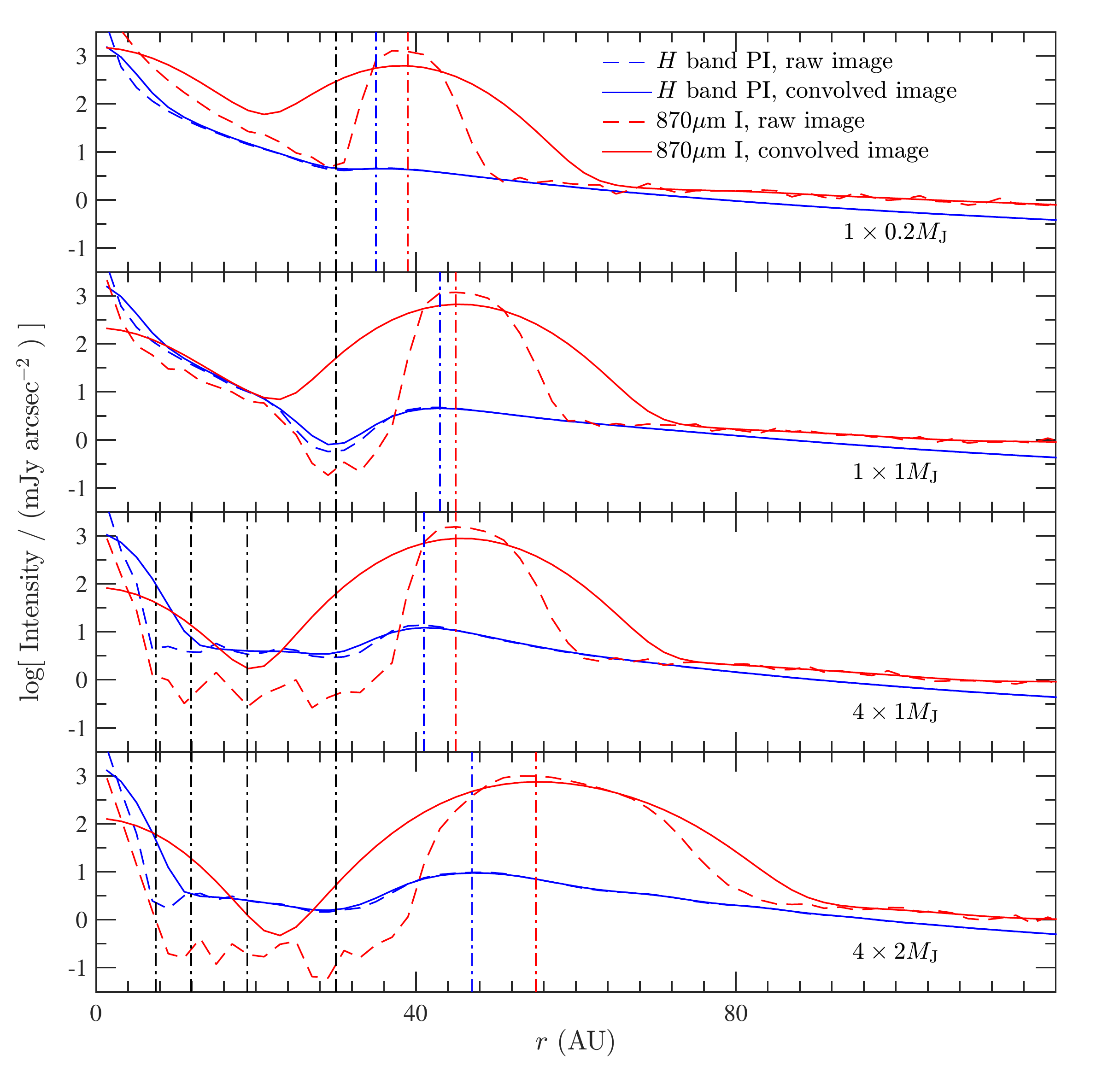}
\end{center}
\figcaption{Azimuthally averaged radial profiles of all images shown in Figure~\ref{fig:image_theta0}. The vertical black dash-dot lines indicate the positions of the planets in each model. The vertical blue and red dash-dot lines mark the positions of the peak intensity in the outer disk in the convolved $H$~band and ALMA images, respectively ($r_{\rm max, out, image}$ in Table~\ref{tab:image_measurement}). In all models, the peak intensity in the outer disk occurs at a larger disk radius at mm than at NIR (more prominent with more massive planets, as in Model~$\modelthree$). Also, NIR gaps are much shallower than mm gaps, in both raw and convolved images.
\label{fig:image_theta0_rp}}
\end{figure}

\begin{figure}
\begin{center}
\includegraphics[trim=0 0 0 0, clip,scale=0.75,angle=0]{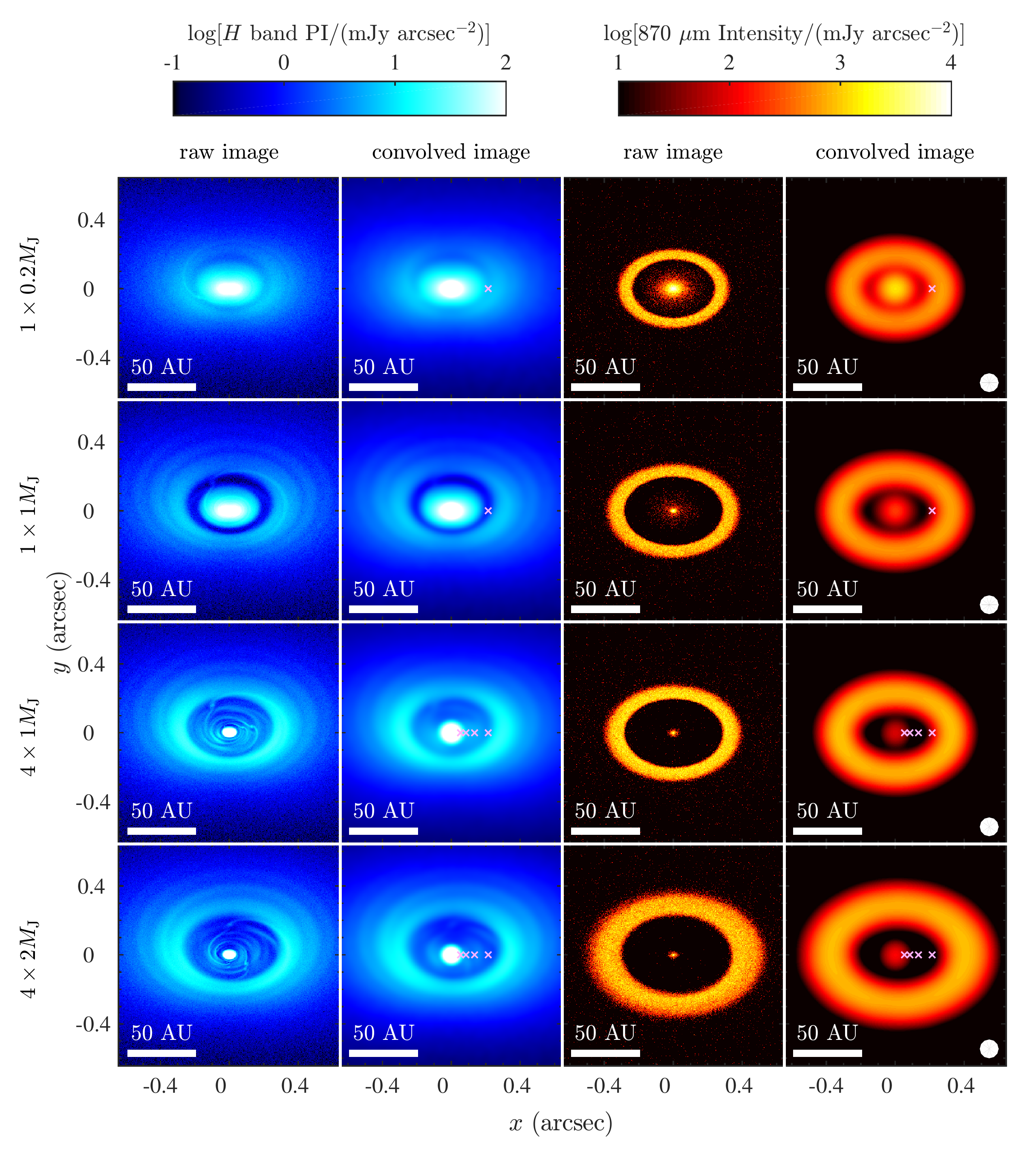}
\end{center}
\figcaption{The same as Figure~\ref{fig:image_theta0}, but for viewing angle $\theta=45^\circ$. The lower half in all panels is the front side of the disk.
\label{fig:image_theta45}}
\end{figure}

\begin{figure}
\begin{center}
\includegraphics[trim=0 0 0 0, clip,scale=0.6,angle=0]{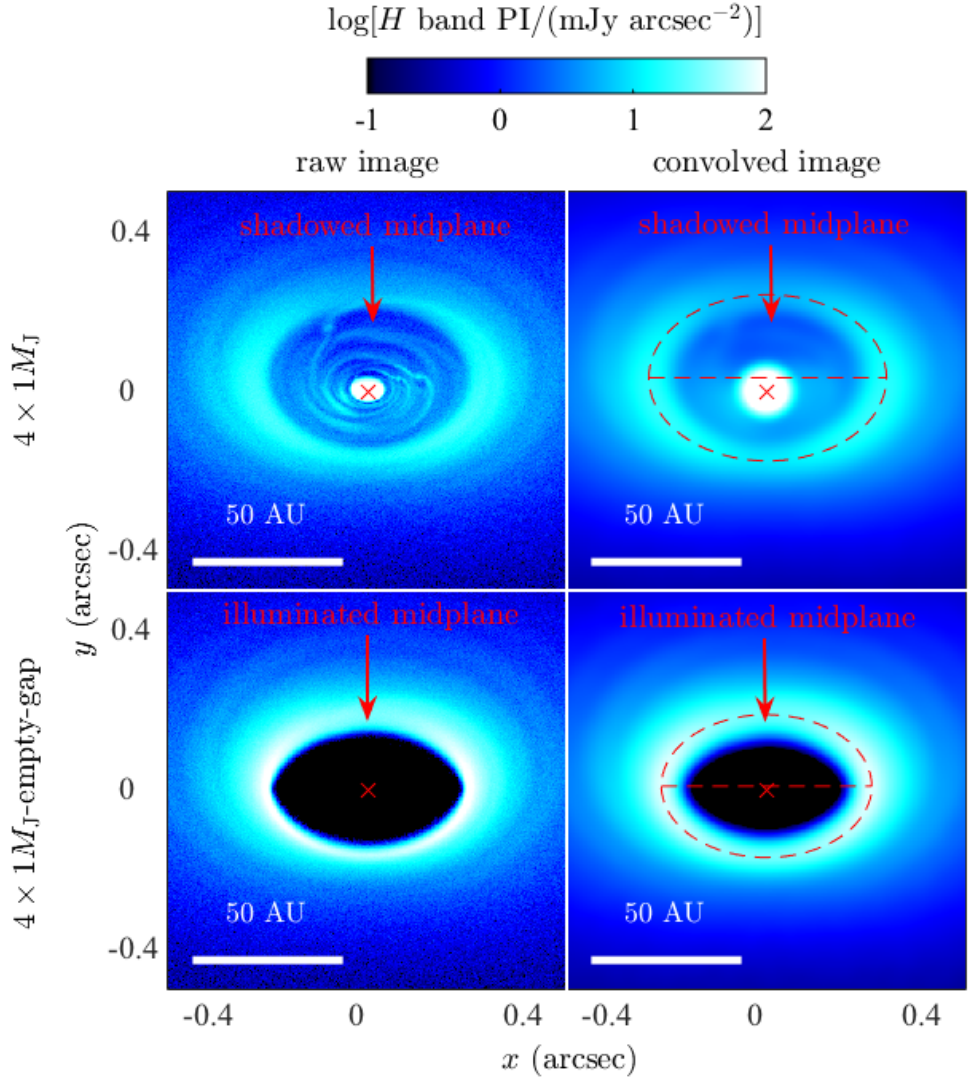}
\end{center}
\figcaption{Comparison of the $H$~band images at a viewing angle $\theta=45^\circ$, between Model~$\modeltwo$ (top row) and Model~$\modeltwo$ with an empty cavity inside 35~AU (bottom row), as discussed in Section~\ref{sec:results-inclined}. The red dashed ellipses in the convolved images are fits to the bright ring in the images, and the red crosses mark the position of the star. The material inside the gap cast a shadow in the middle on the far (up) side of the outer gap wall in Model~$\modeltwo$, while the entire outer gap wall is illuminated in Model~$\modeltwo$-empty-gap. As a result of this geometric effect, the ring in the convolved image of Model~$\modeltwo$ appears to be off-center, while in Model~$\modeltwo$-empty-gap the ring is almost on-center.
\label{fig:image_emptygap}}
\end{figure}

\begin{figure}
\begin{center}
\includegraphics[trim=0 0 0 0, clip,scale=0.75,angle=0]{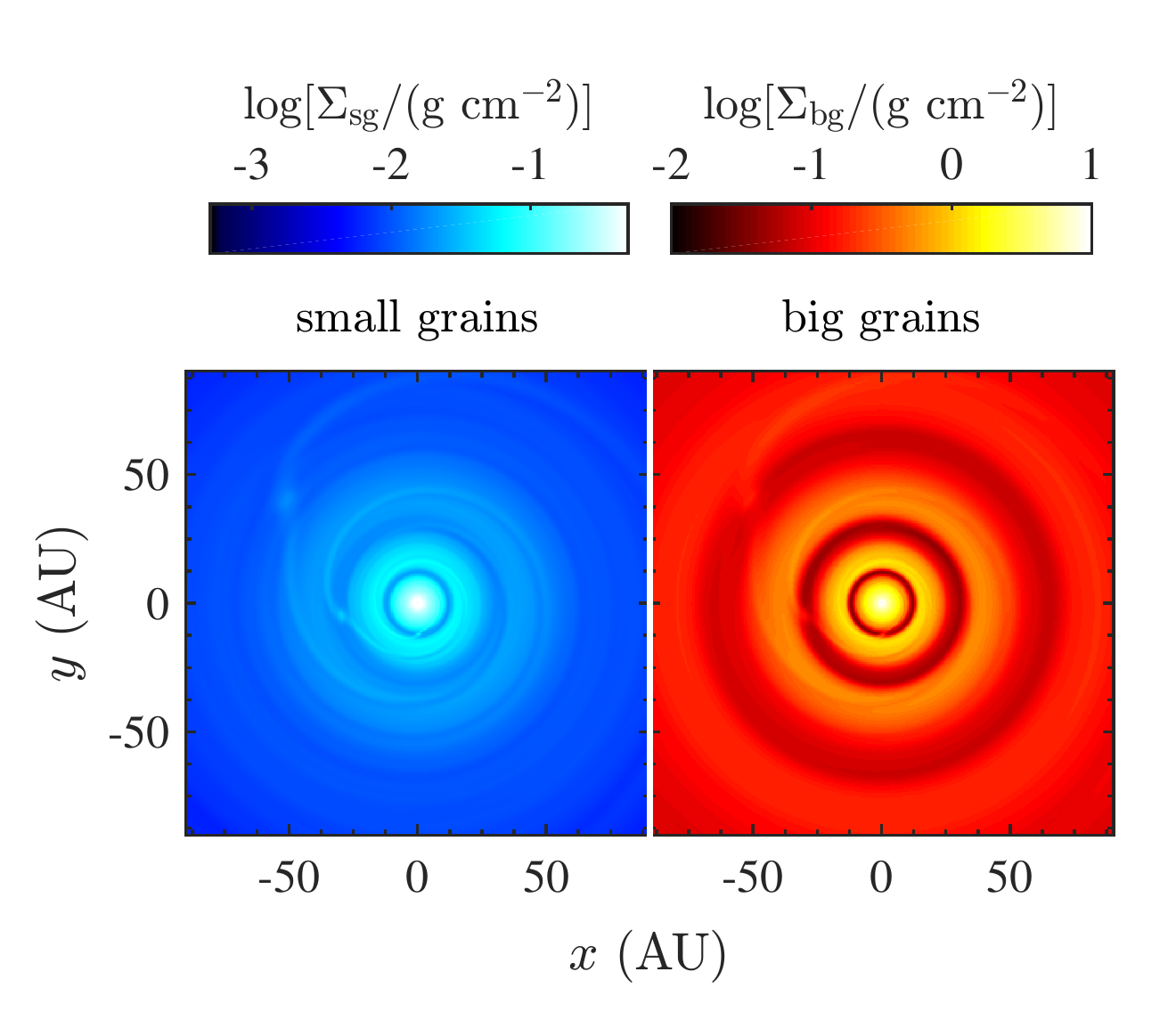}
\includegraphics[trim=0 0 0 0, clip,scale=0.75,angle=0]{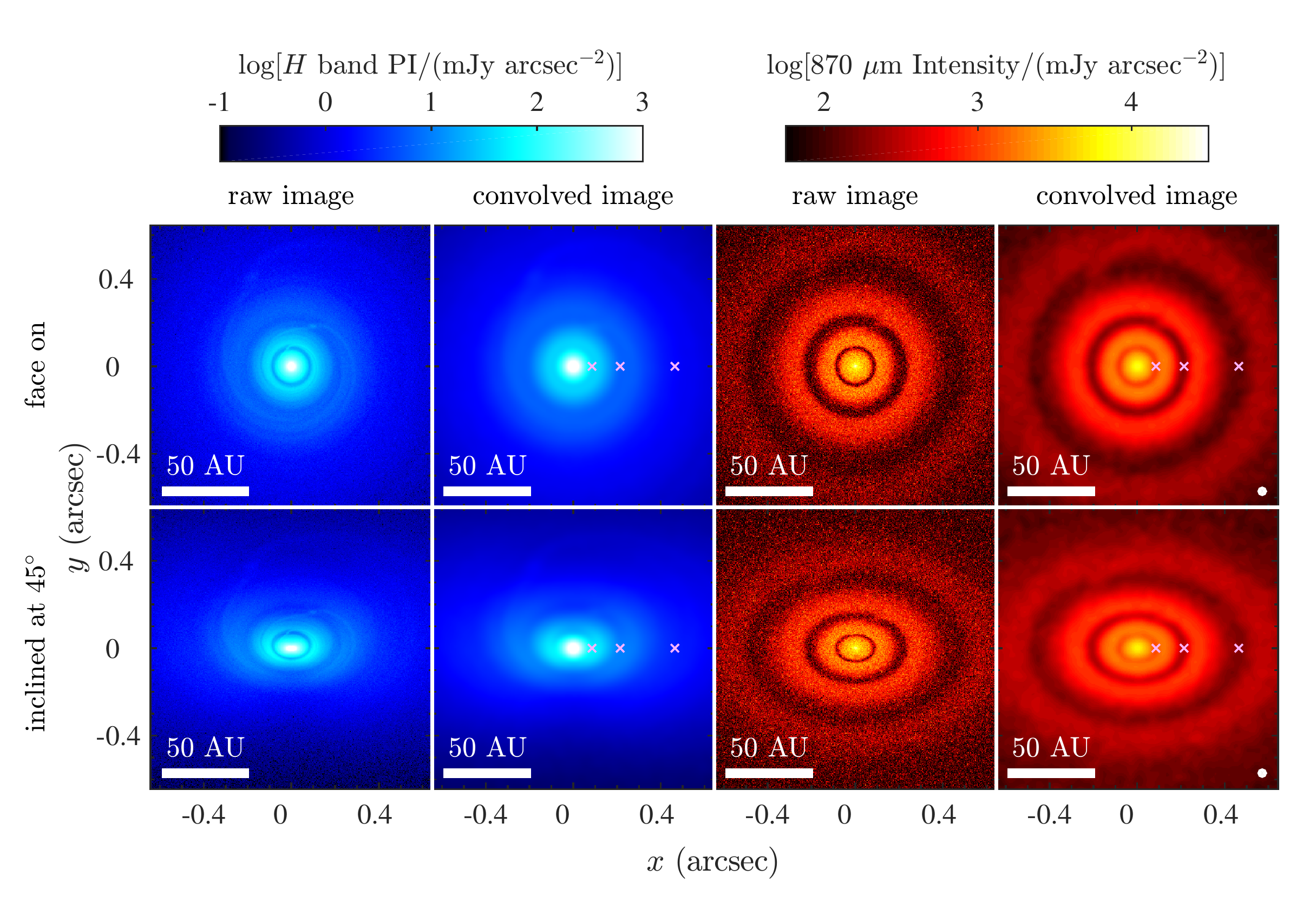}
\end{center}
\figcaption{Model~$\modelhltau$: A 0.17 $\msun$ disk with 3 $0.2\mj$ planets located at 12, 30, and 65 AU 0.2 Myr after their formation, showing multiple gaps well separated by rings. Top: 2D surface density map for the small (left) and big (right) grains. The left panel is also the scaled surface density distribution of the gas, as we assume that the small grains are well mixed with the gas, $\sigmasg=0.1\%\times\sigmag$. Bottom: Raw and convolved $H$~band polarized intensity images (left two columns) and ALMA band 7 (870$~\micron$ continuum) intensity images (right two columns) at face on angle (top row) and an inclined angle $\theta=45^\circ$ (bottom row). The grey crosses in the convolved images mark the orbits of the planets. Systems are assumed to be at 140~pc. The convolved images are convolved by a Gaussian kernel with FWHM=0.04$\arcsec$ at $H$~band and FWHM=0.035$\arcsec$ at ALMA band 7 (beam size indicated at the lower right corner in the convolved mm images). The three planets in this case each opens a narrow gap in both the small and big grains. While signals of the gaps in the convolved NIR images are relatively weak, three narrow gaps well separated by bright rings are clearly visible at mm wavelengths at both inclinations. See Section~\ref{sec:hltau} for detailed discussion.
\label{fig:hltau2d}}
\end{figure}

\begin{figure}
\begin{center}
\includegraphics[trim=0 0 0 0, clip,scale=0.7,angle=0]{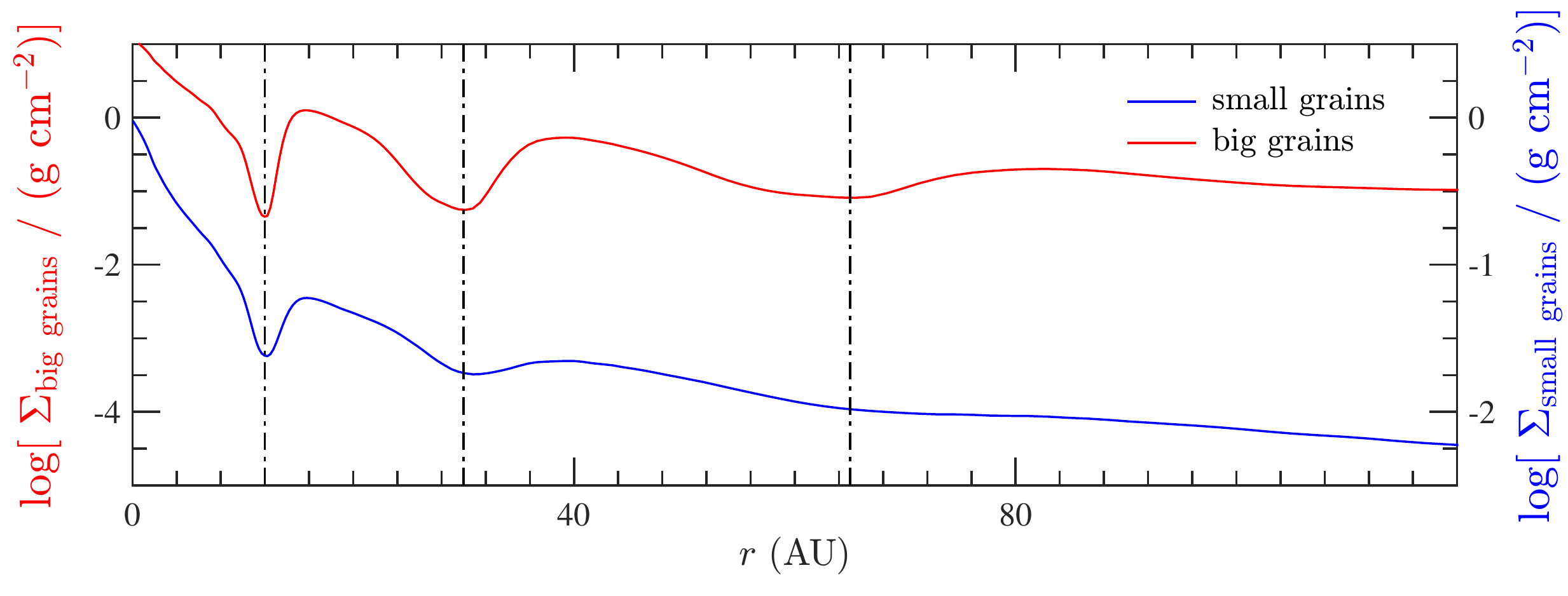}
\includegraphics[trim=0 0 0 0, clip,scale=0.7,angle=0]{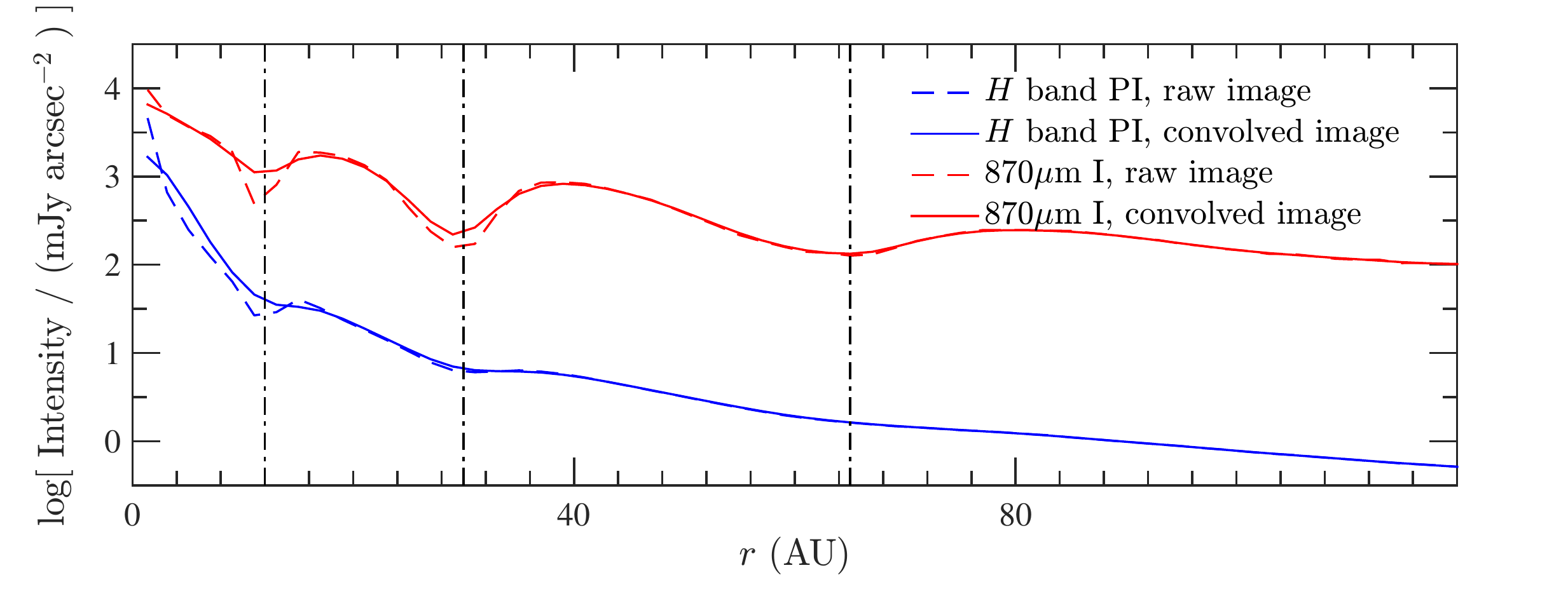}
\end{center}
\figcaption{Radial profiles of the 2D images in Figure~\ref{fig:hltau2d}. Top: The azimuthally averaged surface density radial profile for both the small and big grains from the hydro models. Note that the $y$-axis tick mark labels on the left are for the big grains and the ones on the right are for the small grains. Bottom: Azimuthally averaged radial profiles of all images shown in Figure~\ref{fig:image_theta0}. The vertical black dash-dot lines indicate the positions of the three planets in both panels.
\label{fig:hltau1d}}
\end{figure}

\begin{figure}
\begin{center}
\includegraphics[trim=0 0 0 0, clip,scale=0.7,angle=0]{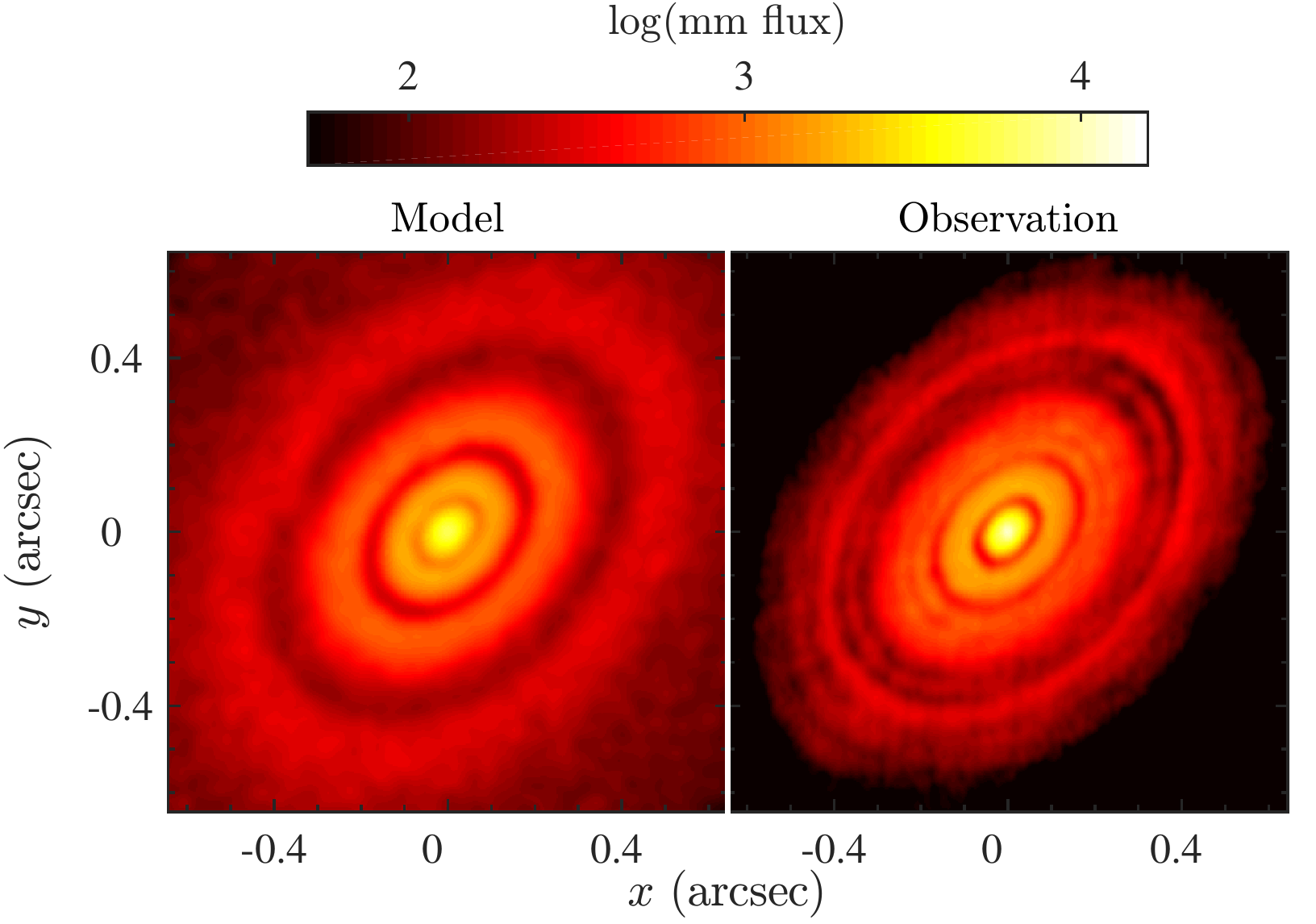}
\end{center}
\figcaption{Comparison between Model~$\modelhltau$ with the ALMA observation of HL Tau (1.0 mm, Band 6+7, \citealt{brogan15}). Model image has an inclination of $45^\circ$ and a position angle of $138^\circ$. The color scale is logarithmic and the unit is arbitrary.
\label{fig:comparison}}
\end{figure}

Overall, the comparisons between models and observations are satisfying, and suggest that the planets-opening-gap scenario is promising to explain the origin of the transitional disks. Our main results are:

\begin{enumerate}

\item We confirm the results in \citet{zhu12, pinilla12b, dejuanovelar13} that the dust filtration effect caused by the presence of a planet \citep{rice06, paardekooper06} is very efficient at piling up the big grains ($\sim$mm-sized) into a ring at the pressure bump outside the gas gap, and evacuating them from the gap. The surface density gap contrast in the big grains between the peak of the ring and the bottom of the gap could be more than 9 orders of magnitude in our models with $\mplanet\geq1\mj$, compared with $\sim1.5$ orders of magnitudes gas/small grains surface density depletion (Figure~\ref{fig:sigma},\ref{fig:sigma_rp}, Table~\ref{tab:sigma_measurement}).

\item It is difficult for a single planet to open a wide gap in the scattered light images, while multiple planets with the same masses can open a giant common gap. In the two models with 1~$M_{\rm J}$ planets, the gap opened by a one $M_{\rm J}$ planet at 30 AU is only $\sim$10~AU wide, while a $\sim$22~AU gap is opened by four 1~$M_{\rm J}$ planets with the outermost planet at the same radius (middle rows in Figures~\ref{fig:image_theta0} and \ref{fig:image_theta0_rp}). This situation may change, as \citet{dejuanovelar13} showed that when the companion mass gets closer to or exceeds the brown dwarf mass limit of 13~$\mj$, the secondary may also open wide gaps in disks.

\item Our multi-planet models with $\mplanet\geq1\mj$ reach gap contrasts (defined as the ratio of the peak intensity in the outer disk to the floor intensity inside the gap) $\sim3-6$ at NIR and $\sim100-1600$ at mm. Both are broadly consistent with observations (bottom three rows in Figure~\ref{fig:sigma},\ref{fig:sigma_rp}, Table~\ref{tab:nirgapcontrast}).

\item We confirm the possible solutions to the ``missing cavity'' problem \citep{dong12cavity} and the cavity size dependence on wavelength effect proposed by \citet{zhu12}, \citet{dejuanovelar13}, and \citet{pinilla12a}. NIR gaps opened by a $0.2~M_{\rm J}$ planet can be too small to be detected with current NIR imaging instruments, while the same planet is still able to create a prominent gap in the big grains due to the dust filtration effects (the top row in Figures~\ref{fig:image_theta0}, \ref{fig:image_theta0_rp}). In our model with $2~M_{\rm J}$ planets, the distance between the gap edge and the outermost planet at 30~AU is twice as large at mm (15~AU) compared to NIR (8~AU, as shown in the bottom row in Figure~\ref{fig:image_theta0_rp} (see also Table~\ref{tab:image_measurement}), consistent with the bigger observed gap size at mm than at NIR in systems like RX~J1604.3-2130 \citep{mayama12, zhang14} and SAO~206462 \citep{andrews11, garufi13}. We note that this is {\it not} caused by the difference in the positions of the peak surface density in the small and big grains.

\item Density waves and streamers, which are produced in this work in 2D hydro simulations due to disk-planet interactions, are marginally traceable in NIR images (Figure~\ref{fig:image_theta0}) of some models, echoing the finding in \citet{juhasz14}. On the other hand, they are essentially absent in mm images.

\item If an inner disk aligned with the outer disk exists inside the gap, it can cast a shadow on the gap wall, which can be seen in scattered light images at intermediate viewing angles (Figure~\ref{fig:image_emptygap}). The absence of this shadow would indicate either no (or very thin) inner disk, or a misaligned inner disk. Furthermore, the ring of the gap edge will appear to be off-center from the star due to this shadowing effect. Therefore, off-centered elliptical rings in scattered light images indicate either a physically elliptical or off-centered gap structure in the cases of no shadowing, as seen in LkCa~15 by \citet{thalmann14}, or the shadowing effect caused by the existence of an inner disk.

\item Multiple narrow gaps well separated by unperturbed disk rings in both big and small grains can be opened by sub-Jupiter mass planets quickly after their formation in disks \citep{pinilla14}. In our simulations, we see that this can be achieved with planets of masses as low as 0.2~$\mj$ that create narrow gaps. Synthetic mm continuum images clearly reveal these gaps and rings, resembling the morphology of the newly released ALMA image of HL Tau (Figure~\ref{fig:comparison}).


\end{enumerate}


\section*{Acknowledgments}

We thank Anthony Boccaletti, Nuria Calvet, Eugene Chiang, Gaspard Duchene, Paul Duffell, Jeffery Fung, Lee Hartmann, Jun Hashimoto, Andrea Isella, Stefan Kraus, John Monnier, Paola Pinilla, Dick Plambeck, Roman Rafikov, Tom Robitaille, Jiming Shi, Jonathan Williams, Mike Wolff, and Ke Zhang for useful discussions and help. This research used the SAVIO computational cluster at UC Berkeley, and the Lawrencium computational cluster resource provided by the IT Division at the Lawrence Berkeley National Laboratory (Supported by the Director, Office of Science, Office of Basic Energy Sciences, of the U.S. Department of Energy under Contract No. DE-AC02-05CH11231). R.D. particularly acknowledges the help from Yong Qin and Kai Song. R.D. would like to thank Zhao Zhu for her support and encouragement in the period of this work. This project is partially supported by NASA through Hubble Fellowship grants HST-HF-51333.01-A (Z.Z.) and HST-HF-51320.01-A (R.D.) awarded by the Space Telescope Science Institute, which is operated by the Association of Universities for Research in Astronomy, Inc., for NASA, under contract NAS 5-26555. We also thank the anonymous referee for constructive suggestions that largely improved the quality of the paper. This paper makes use of the following ALMA data: ADS/JAO.ALMA\#2011.0.00015.SV. ALMA is a partnership of ESO (representing its member states), NSF (USA) and NINS (Japan), together with NRC (Canada) and NSC and ASIAA (Taiwan), and KASI (Republic of Korea), in cooperation with the Republic of Chile. The Joint ALMA Observatory is operated by ESO, AUI/NRAO and NAOJ. The National Radio Astronomy Observatory is a facility of the National Science Foundation operated under cooperative agreement by Associated Universities, Inc.


\appendix
\section{SED, and Images at $10\micron$ and $100\micron$} \label{sec:appendix}

SEDs at viewing angles $\theta=0^\circ$ and $45^\circ$ for Models~$\modelzero$, $\modelone$, $\modeltwo$, and $\modelthree$ are shown in Figure~\ref{fig:sed}. The NIR dip on the SED that is similar to transitional disks is not created by the gaps. This is a known characteristic of the Whitney code in full disk SED modeling (Figure 1, \citealt{whitney13}, in which a full class II T Tauri disk also shows this dip around 1-10 $\micron$ as well as the silicate feature). The dip may be due to the fact that the models do not include emission from the warm gas inside the dust sublimation radius. Also, specific choices on the dust sublimation front (e.g. sublimation temperature, and the shape of the rim) may also affect the SED at short wavelengths. Here, the relative differences between models show the effects of different gaps opened by planets. Models with multiple planets have less IR excess than models with a single planet at $\sim10-100\micron$, and more emission at wavelengths beyond $\sim100\micron$. This is because the gaps opened by multiple planets are bigger, resulting in lower emission from the gap region. Smaller gaps intercept less starlight, and therefore the outer disk receives more starlight resulting in higher grain temperature and more emission at long wavelengths. Nevertheless, the difference in the SEDs between models is marginal. In addition, all of them are similar to the SED of a full class II T Tauri disk produced by the Whitney code (Figure~1, \citealt{whitney13}).

\begin{figure}
\begin{center}
\includegraphics[trim=0 0 0 0, clip,scale=0.75,angle=0]{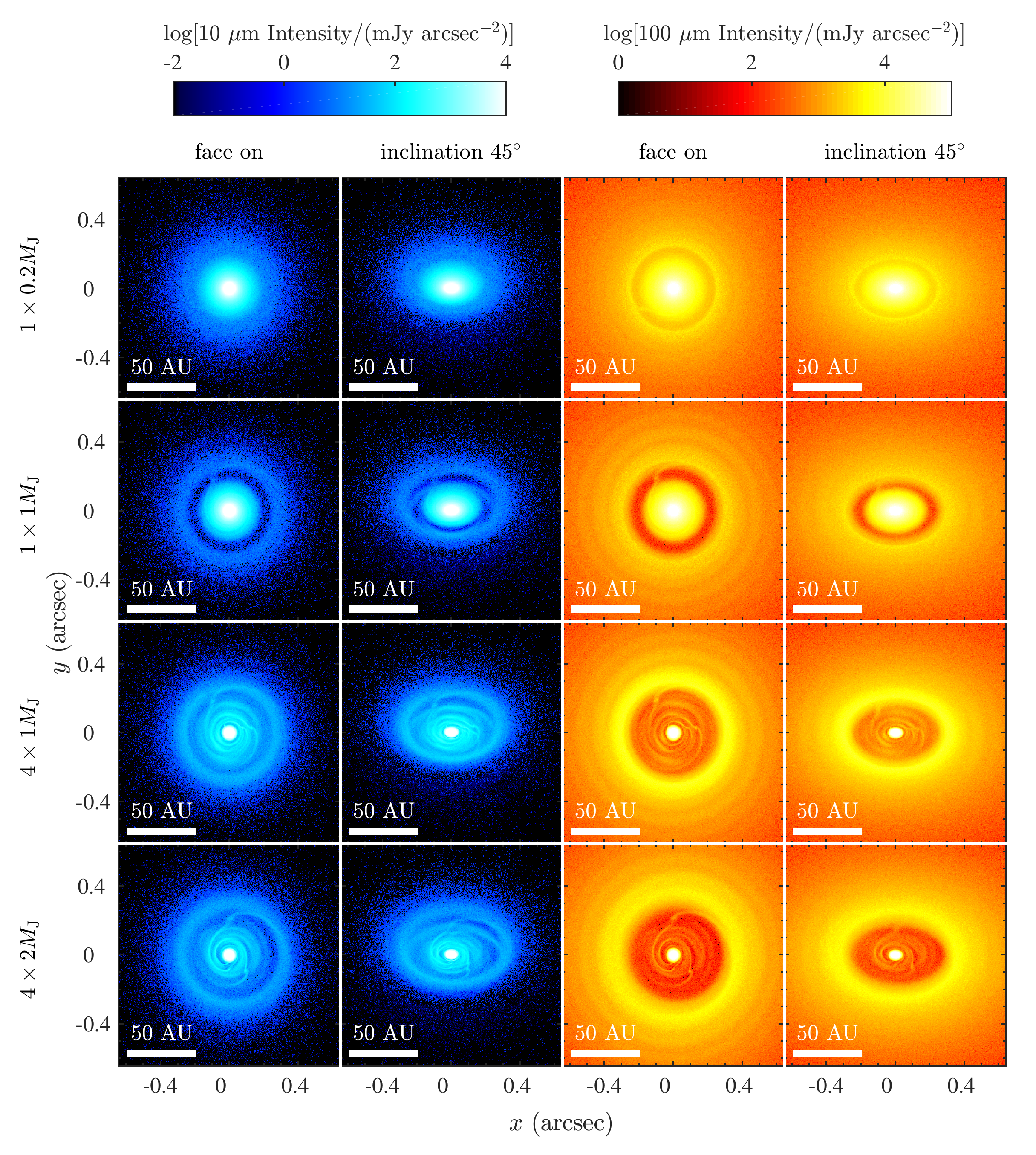}
\end{center}
\figcaption{Raw $10~\micron$ intensity images (left two columns) and 100$~\micron$ intensity images (right two columns) for our models (model names labeled on the left) at viewing angles $\theta=0^\circ$ and $45^\circ$. At $10~\micron$, gaps appear to be deeper than at $100~\micron$, and the shadow casted by the inner disk is visible on the outer gap wall in the inclined disk images.
\label{fig:image_10_100um}}
\end{figure}

\begin{figure}
\begin{center}
\includegraphics[trim=0 0 0 0, clip,scale=0.72,angle=0]{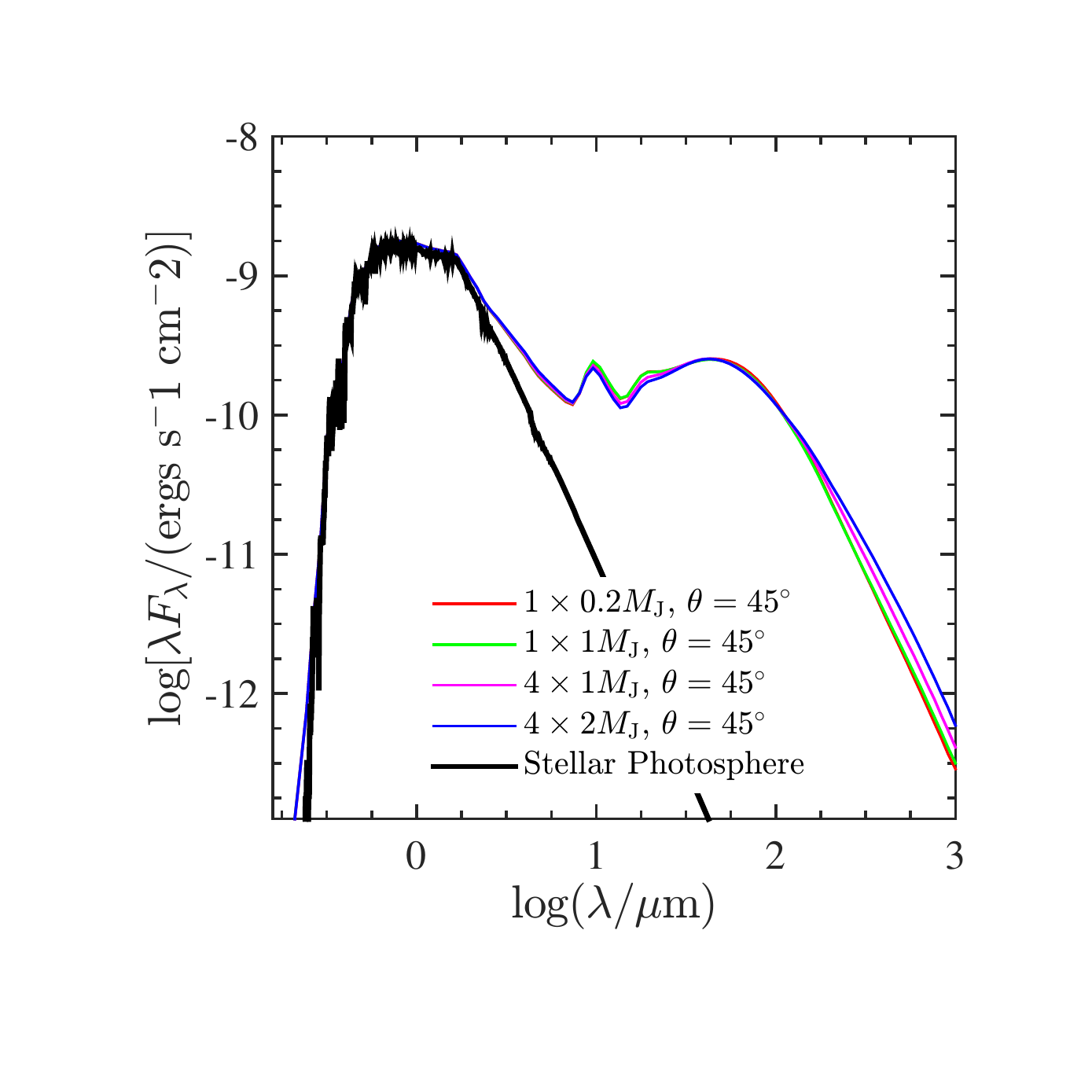}
\end{center}
\figcaption{SEDs for Models~$\modelzero$, $\modelone$, $\modeltwo$, and $\modelthree$ at a viewing angle $\theta=45^\circ$. The difference between models is marginal.
\label{fig:sed}}
\end{figure}

The raw images at $10~\micron$ and $100~\micron$  for Models~$\modelzero$, $\modelone$, $\modeltwo$, and $\modelthree$ are shown in Figure~\ref{fig:image_10_100um} from viewing angles $\theta=0^\circ$ and $45^\circ$. The images at $10~\micron$ appear to be less empty comparing with at $100~\micron$. Also, the dark lane at the mid-plane on the far (up) side of the gap wall is clearly visible at $10~\micron$. Both features indicate that $10~\micron$ images are mostly dominated by scattered light, while signals at $100~um$ mainly come from dust thermal emission.

\clearpage

\begin{table}
\begin{center}
\caption{Model Properties}
\begin{tabular}{cccccccc}
\tableline\tableline
Model  &  Planet Mass & Planet Position & Evolution Time & $M_{\rm disk}$\tablenotemark{a} & Particle Size\tablenotemark{b} & \multicolumn{2}{c}{Angular Resolution in Convolved Images\tablenotemark{c}} \\
                      &        $\mj$  &  AU                      & yr & $\msun$ & mm & $H$ Band & ALMA Band 7 \\
\tableline
$\modelzero$ & 0.2 & 30.0 & 4$\times$10$^{5}$ & 0.04 & 1 & 0.04$\arcsec$ & 0.10$\arcsec$ \\
$\modelone$ & 1 & 30.0 &  4$\times$10$^{5}$ & 0.04 & 1 & 0.04$\arcsec$ & 0.10$\arcsec$  \\
$\modeltwo$ & 1/1/1/1 & 30.0/18.9/11.9/7.5 & 4$\times$10$^{5}$ & 0.04 & 1 & 0.04$\arcsec$ & 0.10$\arcsec$ \\
$\modelthree$ & 2/2/2/2 & 30.0/18.9/11.9/7.5 & 4$\times$10$^{5}$ & 0.04 & 1 & 0.04$\arcsec$ & 0.10$\arcsec$ \\
\tableline
$\modelhltau$ & 0.2/0.2/0.2 & 12/30/65 & 2$\times$10$^{5}$ & 0.2 & 1 & 0.04$\arcsec$ & 0.035$\arcsec$ \\
\tableline
\end{tabular}
\tablecomments{Properties of the models. $^a$ The initial gas disk mass. $^b$ The size of the particles in the 2 fluid hydro simulations (see Section~\ref{sec:hydro} for details). $^c$ The angular resolution in the convolved images at $H$ band and ALMA band 7, defined as the FWHM of the PSF used in the convolution process (see Section~\ref{sec:mcrt} for details).}
\label{tab:model}
\end{center}
\end{table}

\begin{table}[h]
\begin{center}
\caption{Gap Property Measurements in the Surface Density of the Grains}
\begin{tabular}{@{}cccccc@{}}
\toprule
Model                                  & Grains       & $\Sigma_{\rm min,gap}$ & $\Sigma_{\rm max, out}$ & $r_{\rm max,out,\Sigma}$ & $\zeta_\Sigma$ \\ \midrule
(1)                                    & (2)          & (3)                    & (4)                     & (5)                & (6)                  \\ \cmidrule(lr){3-5}
                                       &              & \multicolumn{2}{c}{g cm$^{-2}$}                  & AU                 &                      \\ \midrule
\multirow{2}{*}{$1\times0.2M_{\rm J}$} & Small Grains & $3.0\times10^{-3}$     & $4.0\times10^{-3}$      & 40          & 1.3                  \\
                                       & Big Grains   & $5.9\times10^{-4}$     & 2.1                 & 38          & 3500               \\ \midrule
\multirow{2}{*}{$1\times1M_{\rm J}$}   & Small Grains & $1.9\times10^{-4}$     & $3.5\times10^{-3}$      & 47          & 18                 \\
                                       & Big Grains   & N/A                    & 1.4                 & 45          & $\gtrsim10^9$        \\ \midrule
\multirow{2}{*}{$4\times1M_{\rm J}$}   & Small Grains & $1.3\times10^{-4}$     & $4.1\times10^{-3}$      & 47          & 32                 \\
                                       & Big Grains   & N/A                    & 1.3                 & 45          & $\gtrsim10^9$        \\ \midrule
\multirow{2}{*}{$4\times2M_{\rm J}$}   & Small Grains & $6.6\times10^{-5}$     & $3.0\times10^{-3}$      & 56          & 46                 \\
                                       & Big Grains   & N/A                    & 0.51                & 55          & $\gtrsim10^9$        \\ \bottomrule
\end{tabular}
\tablecomments{Col. (1): Model name. Col. (2): Grain type. Col. (3): Minimum azimuthally averaged surface density inside the gap; not well defined in $\sigmabg$ in models with $\mplanet\geq1~\mj$ as $\sigmabg$ reaches the hard floor set in hydro simulations. Col. (4): Maximum azimuthally averaged surface density in the outer disk. Col. (5): The location of $\Sigma_{\rm max, out}$. Col. (6): Gap contrast $\zeta_\Sigma=\Sigma_{\rm max, out}/\Sigma_{\rm min, gap}$; upper limits for the big grains in Models with $\mplanet\geq1\mj$.}
\label{tab:sigma_measurement}
\end{center}
\end{table}

\begin{table}[h]
\begin{center}
\caption{Gap Property Measurements in Images}
\begin{tabular}{@{}cccccccccccccc@{}}
\toprule
Model                                & Band    & \multicolumn{2}{c}{$r_{\rm gap, in, image}$} & \multicolumn{2}{c}{$r_{\rm gap, out, image}$} & \multicolumn{2}{c}{$\deltagap$} & \multicolumn{2}{c}{$r_{\rm max, out, image}$} & $I_{\rm min, gap}$ & $I_{\rm max, out}$ & $\zeta_{\rm image,conv}$ & $\zeta_{\rm image,raw}$ \\ \midrule
(1)                                  & (2)     & \multicolumn{2}{c}{(3)}               & \multicolumn{2}{c}{(4)}                & \multicolumn{2}{c}{(5)}       & \multicolumn{2}{c}{(6)}                      & (7)                   & (8)                  & (9)                  & (10)          \\ \cmidrule(l){3-12} 
                                     &         & AU               & arcsec             & AU               & arcsec              & AU           & arcsec         & AU                & arcsec               & \multicolumn{2}{c}{mJy arcsec$^{-2}$}        &               &          \\ \midrule
\multirow{2}{*}{$1\times0.2M_{\rm J}$} & $H$ (PI)     & N/A              & N/A              & N/A               & N/A                & N/A            & N/A           & 35                & 0.25                 & 4.4                   & 4.5                 & 1.02           & 1.1 \\
                                     & ALMA B7 & 12               & 0.09                & 31              & 0.22                & 19          & 0.14            & 39                & 0.28                 & 61                  & 624                & 10          & 270 \\ \midrule
\multirow{2}{*}{$1\times1M_{\rm J}$} & $H$ (PI)     & 25               & 0.18               & 35               & 0.25                & 10            & 0.07           & 43                & 0.31                 & 0.8                   & 4.6                 & 5.7           & 8.3 \\
                                     & ALMA B7 & N/A              & N/A                & 38               & 0.27                & N/A          & N/A            & 45                & 0.32                 & 7.0                  & 670                & 96          & $\sim10^4$ \\ \midrule
\multirow{2}{*}{$4\times1M_{\rm J}$} & $H$ (PI)     & 12               & 0.09               & 34               & 0.24                & 22           & 0.16           & 41                & 0.29                 & 3.5                   & 12.2                 & 3.5           & 4.8 \\
                                     & ALMA B7 & N/A              & N/A                & 38               & 0.27                & N/A          & N/A            & 45                & 0.32                 & 1.7                   & 884                & 516         & $\sim10^{3.5}$ \\ \midrule
\multirow{2}{*}{$4\times2M_{\rm J}$} & $H$ (PI)     & 10               & 0.07               & 38               & 0.27                & 28           & 0.20           & 47                & 0.34                 & 1.6                   & 9.5                  & 6.0           & 6.7 \\
                                     & ALMA B7 & N/A              & N/A                & 45               & 0.32                & N/A          & N/A            & 55                & 0.39                 & 0.5                   & 748                & 1593        & $\sim10^{3.5}$ \\ \bottomrule
\end{tabular}
\tablecomments{Col. (1): Model name. Col. (2): Observational bands. $H$ band is the polarized intensity. ALMA B7 is ALMA Band 7, continuum at $870~\micron$. Col. (3) and (4): Inner and out gap edge in units of AU and arcsec (objects are at 140~pc from us), defined as the locations in the gap where the intensity reaches half of the peak value in the outer disk. For models with $\mplanet\geq1\mj$, thermal emission inside $r_{\rm gap, out, image}$ never reaches half of the peak intensity in the outer disk, so $r_{\rm gap, in, image}$ is not defined at ALMA B7.  The NIR gap in Model~$\modelzero$ is too shallow to ever reach half of the peak density in the outer disk, so both gap edges are not defined. Col. (5): $r_{\rm gap, out, image}-r_{\rm gap, in, image}$. Col. (6): Radii of the peak intensity in the outer disk. Col. (7): Minimum intensity in the gap. Col. (8): Peak intensity in the outer disk. Col. (9): Gap contrast $\zeta_{\rm image,conv}=I_{\rm max, out}/I_{\rm min, gap}$. Col. (3) to (9) all refer to convolved images. Col. (10): Similar to Col. (9), but for raw images. $I_{\rm min, gap}$ in raw images is chosen as a general floor value inside the gap, as the absolute minimum may be too low and is affected by the noise in MCRT simulations.}
\label{tab:image_measurement}
\end{center}
\end{table}

\begin{table}[h]
\begin{center}
\caption{Comparison of NIR Gap Contrast between Observations and Models}
\begin{tabular}{@{}ccc@{}}
\toprule
System Name & Gap Contrast at $H$~Band & References \\ 
\midrule
RX~J1604.3-2130 & $3.6\pm0.5$  & \citet{mayama12} \\ 
HD~142527 & $4.2\pm1.2$ &  \citet{canovas13}\\ 
SAO~206462 & $3.3\pm1.9$ &  \citet{garufi13}\\ 
PDS~70 & $3.5\pm0.7$ (along major axes) & \citet{hashimoto12}\\
\midrule
$4\times1M_{\rm J}$ & 3.5 & This work \\ 
$4\times2M_{\rm J}$ & 6.0 & This work \\ 
\bottomrule
\end{tabular}
\label{tab:nirgapcontrast}
\tablecomments{The gap contrast in our models is the $\zeta_{\rm image,conv}$ defined in Table~\ref{tab:image_measurement}. The gap contrast in observed systems is measured in a similar manner, in which the value and the quoted error are the mean and standard deviation over different azimuthal angles. The first three observed systems are relatively face-on, while PDS~70 has a non-trivial inclination of 50$^\circ$, so only major axes are taken into account. The floor intensity inside the gap is detected only in HD~142527. In the other three systems the possible inner edge of the gap (if exists) is blocked by the inner working angle in observations, so the measured gap contrast may be a lower limit. Observational noise is not added in our model images, so measured values may be upper limits. Nevertheless, the agreement between model results and observations is encouraging.}
\end{center}
\end{table}


\end{document}